# Wall-Modeled Large-Eddy Simulation of Turbulent Non-Newtonian Power-Law Fluid Flows

Mohammad Taghvaei and Ehsan Amani[†]

*Department of Mechanical Engineering, Amirkabir University of Technology (Tehran Polytechnic), Iran*

For high-fidelity predictions of turbulent flows in complex practical engineering problems, the Wall-Modeled (WM) Large-Eddy Simulation (LES) has aroused great interest. In the present study, we prove that the conventional Wall-Stress Models (WSMs) developed for WMLES of Newtonian fluids fail to predict the shear-thinning-induced drag reduction in power-law fluids. Therefore, we propose novel algebraic, integrated, and Ordinary-Differential-Equation (ODE) WSMs, for the first time, for WMLES of power-law Non-Newtonian (NN) fluids and assess their performance against reference Wall-Resolved (WR) LES solutions. In addition, the effects of the key model parameters, including the WSM type, sampling height, sampling cell, and axial grid resolution are explored, and it is revealed that turbulent NN flow predictions have a much higher sensitivity to the choice of WSM, compared to their Newtonian counterparts. It is manifested that, in contrast to WRLES, accurate modeling of the mean apparent and subgrid-scale NN viscosities in the NNODE model can improve the predictions considerably. Therefore, closures with lower uncertainties on coarse WMLES grids are sought for these terms. Finally, the best performance for the present test cases is obtained via the integrated NN WSM, sampling at the lower edge of the log layer within the third off-the-wall grid cell. Nevertheless, the new NNODE WSM can have an advantage in the presence of non-equilibrium effects in more complex problems.

---

[†] Corresponding author. Address: Mechanical Engineering Dept., Amirkabir University of Technology (Tehran Polytechnic), 424 Hafez Avenue, Tehran, P.O.Box: 15875-4413, Iran. Tel: +98 21 64543404. Emails: eamani@aut.ac.ir (E. Amani) taghvaei@aut.ac.ir (M. Taghvaei).

## 1. Introduction

To capture unsteady or complex multi-physics turbulent flow phenomena in many applications, Reynolds-Averaged Navier-Stokes (RANS) models are inadequate. Despite the rapid increase in the processing speed of modern computers, there are still significant limitations to the use of Direct Numerical Simulation (DNS) or even well-resolved Large-Eddy Simulation (LES) approaches for practical wall-bounded flows with high Reynolds (Re) numbers. In contrast to the outer region of the wall boundary layer, in which the size of energetic turbulent structures is of the order of the boundary layer thickness, $\delta$, in the inner part of the boundary layer, the size of turbulent structures scales down to the viscous length, $\delta_v = v/u_\tau$ where $v$ and $u_\tau$ are the molecular kinematic viscosity and friction velocity, respectively. Resolving these small structures for a complex industrial problem requires an incredibly large number of computational cells near the walls, leading to a huge computational cost, beyond the power of current computers. Even at moderate Re numbers, almost 90 percent of the computational burden is placed upon the resolution of the inner region of the Turbulent Boundary Layer (TBL) [1]. Choi and Moin [2] estimated the number of grid cells required for a Wall-Resolved LES (WRLES) of the inner layer to be proportional to $Re^{13/7}$ while for the outer layer, the resolution scales linearly with Re.

In recent years, much attention has been paid to Wall-Modeled LES (WMLES) techniques. Several studies have implemented these strategies even into discontinuous Galerkin and lattice-Boltzmann methods [3, 4]. Many WMLES models have been proposed to avoid the high computational cost of the inner layer resolution in LES. These can be categorized into two main groups [1, 5, 6]: hybrid RANS/LES and Wall-Stress Modeling (WSM). In hybrid RANS/LES methods, the inner part of TBL is solved by a RANS model, while LES captures the outer region. Readers can consult references [7, 8] for reviews on this category. In the latter approach, LES is used in the whole domain, and the wall shear stress is computed and fed as the boundary condition to LES. Based on the procedure of wall-stress calculation, WSM is divided into four algebraic, integrated, Ordinary Differential Equation (ODE), or Partial Differential Equation (PDE) groups of models. In the PDE methods, a separate mesh is generated between the wall and the outer part of TBL, in which the simplified thin boundary-layer equations are solved with the input from the original LES as the boundary condition. On the contrary, algebraic, integrated, and ODE models do not require a separate grid and pose less implementation complexity and much lower computational overhead, making them more efficient choices for general unstructured grids and complex geometries.

The idea of WSM approaches, i.e., accounting for the dynamics of the inner layer by prescribing the correct value of the filtered shear stress at walls, dates back to the pioneering works by Deardorff [9] and Schumann [10]. Since then, many improvements have been proposed to this area, however, there are still many issues that have to be tackled to achieve reliable results for practical applications. Here, the most important progress and developments in this area are outlined to highlight the gap in the literature. More comprehensive reports can be found elsewhere, see e.g., the reviews by Larsson, et al. [6] and Bose and Park [5].

One of the challenges of WSM is taking the Non-Equilibrium Effects (NEE) into account, especially the correct prediction of separating flows. Cabot [11], for the first time, applied the PDE method and examined the effect of considering the pressure gradient with/without the advection and unsteady terms. No difference was seen for the prediction of channel flow; however, a slight improvement was reported for the prediction of separation inception in a backward-facing step case. Afterward, Balaras, et al. [12] employed the PDE and algebraic methods for rotating/non-rotating channel flow and square ducts. They observed no difference in average velocity predictions, but the PDE method captured the Root-Mean-Square (RMS) velocities slightly better, although the distinction reduced when the Reynolds number increased. They reported that the PDE method costs 10-15% more than the algebraic one. Hoffmann and Benocci [13] employed the ODE model for the first time by integrating TBL from the wall to the first off-the-wall cell, considering the viscous, pressure, and unsteady terms and neglecting the advection term. For this purpose, they assumed that pressure does not change between the wall and the first off-the-wall cell and approximated the unsteady term by a first-order discretization. Their model showed a slightly better performance in RMS velocity prediction compared to several previous algebraic models. Cabot [14] exploited the PDE method in the backward-facing step and showed that when NEE terms are included, all terms should be kept in TBL simultaneously. Omitting both pressure and advection terms, they also could predict the reattachment location well, although there was about a 10 percent discrepancy when only one of these terms was accounted for. The predicted skin friction coefficient and average velocity in the separation zone agreed well with the experiment when neither pressure nor advection term was considered; however, considering both terms (full TBL equation) led to the best result in the reattachment zone. As discussed by Larsson, et al. [6], equilibrium assumption in cases involving separation can capture some NEE since in WMLES, 80 percent of TBL are resolved. In addition, owing to the much faster dynamics of the modeled inner layer, the equilibrium assumption for this region could be justified. Furthermore, in the overlap region ($30 < y^+ < 50$), pressure and advection counterbalance each other. Duprat, et al. [15] incorporated the effect of pressure gradient into RANS-type eddy viscosity for the ODE method by introducing a new velocity scale. Their new RANS eddy viscosity generated better results than the Spalding law of the wall for channel flows with stream-wise periodic constrictions; nevertheless, such improvement has not been observed in the recent study by Mukha, et al. [16] for the backward-facing step case, advocating the necessity of including all NEE terms or none of them. To relax the assumption of constant NEE terms across the inner layer in the ODE method, Hickel, et al. [17] proposed a piecewise-linear approximation for the pressure and advection terms.

In recent years, new models for considering NEE have been developed. However, these models are often computationally expensive and have not improved the predictions strikingly [18, 19].

Another challenge of WSM, and all WMLES, is the correct prediction of Transition To Turbulence (TTT). Bodart and Larsson [20] proposed a flow sensor to estimate the state of the laminar or turbulent flow. This model worked well with WMLES since the instability waves responsible for TTT scale with TBL thickness which is mostly resolved in WMLES. Park and Moin [21] used the sensor-based technique and found that it is required to refine the grid in the wall-normal direction in the transition region to accurately capture TTT.

Probably, the most prevalent issue of all WMLES approaches is the well-known Log-Layer Mismatch (LLM), the over-prediction (positive mismatch) or under-prediction (negative mismatch) of the predicted mean velocity profile in viscous units compared to the log law. For the first time, Werner and Wengle [22] proposed the integral method, specifically the integral form of the exponential law of the wall (called WW model here), to compute wall shear stress. The merit of their integral model is the elimination of the need for the calculation of the averaged velocity as an input of the wall model; therefore, it can be used in complex cases where the homogenous directions cannot be detected easily. The integral model performed better than its algebraic counterpart in decreasing LLM. Temmerman, et al. [23] directly deployed the algebraic and integral form of the logarithmic law of the wall and WW model for channel flows with stream-wise periodic constrictions on both medium and coarse grids. They discovered that the integral form of the wall model on coarse meshes performs better than the algebraic one, yet the results showed no improvement on a medium mesh. In addition, they noticed the number of stream-wise cells impacts the stream-wise RMS velocity but the stream-wise average velocity. Kawai and Larsson [24] evaluated errors in a few first cells off the wall and argued that at least two computational cells are required to resolve turbulence structures in the logarithmic layer, therefore, choosing the first cell for sampling data inevitably produces false output. They revised the classical view of opting for the first off-the-wall cell for sampling by employing the next cells and succeeded to offer a remedy to LLM. Later their idea of opting for the next consecutive of-the-wall cells was supported by Lee, et al. [25] and Rezaeiravesh, et al. [26]. In addition to the sampling cells, the height of the sampling point was another major factor asserted by Kawai and Larson, which needs further investigation. Despite Temmerman's finding, they reported that coarsening the computational mesh did not alter results, which requires supplementary examinations. By studying the available data, Mukha, et al. [16] also obtained better results by using the second cell as the sampling cell; nevertheless, they rejected Kawai and Larson's reasoning for this choice (the error in the information at the wall-adjacent cell) and claimed that the first cell is not suitable for sampling because its velocity is not a true representation of actual averaged velocity. Yang, et al. [27] pointed out that the non-local nature of Kawai and Larson's method degrades the parallel computation efficiency. In addition, ensuring the existence of a proper node for data sampling is impossible for some applications involving complex geometries and unstructured meshes with sharp boundary edges. Consequently, they offered a new method based on time averaging, over a period equal to the convection time scale, to solve the LLM problem.

The other drawback of WMLES is in the prediction of Multi-Physics Phenomena (MPP) coupled with near-wall turbulence. WMLES aims to eliminate the need for resolving high-frequency information to govern the fluid motion; however, some multi-physics phenomena; e.g., particle deposition on walls, combustion, radiation, sound generation, etc.; depend on the small turbulence scales in the inner layer. This usually necessitates the use of ad hoc models, e.g., dispersion models for particle-turbulence interaction near walls [28, 29]. For turbulent non-Newtonian fluid flows, Amani, et al. [30] showed that shear-thinning rheology can have a substantial impact on the high-frequency part of the energy spectrum. They proposed a new dynamic Sub-Grid Scale (SGS) model to take this effect into consideration. In their subsequent study [31], they evaluated the new dynamic SGS model for fully developed turbulent pipe flows at moderate Re numbers. Although their results manifested that the fluid rheology effect on SGS closure is negligible in a well-resolved LES where the filter width to DNS cell size is not very large, this probably is not the case for more complex practical situations where the use of WMLES is unavoidable. Therefore, standard Newtonian WMLES may be inadequate for non-Newtonian turbulent flows in wall-bounded flows and ad hoc models or modifications to WMLES closures may be necessary for these applications. This is the main focus of the present study.

In the present research, we manifest the inadequacy of the conventional WSM, developed for Newtonian fluids, for the use in WMLES of Turbulent Non-Newtonian Flows (TNNF). Then, new algebraic, integrated, and ODE WSM models are proposed and assessed against WRLES reference solutions. Afterward, the effect of model key parameters, including the height of sampling point, the number of computational cells between the wall and the sampling point, and the type of WSM are analyzed carefully. Finally, the recommendations for using the new models are provided.

## 2. Governing equations

### 2.1. LES equations

Applying the filtering operation on the continuity and momentum equations for an incompressible non-Newtonian fluid, LES governing equations can be derived as:

$$\frac{\partial \bar{u}_j}{\partial x_j} = 0 \tag{1}$$

$$\frac{\partial \bar{u}_i}{\partial t} + \frac{\partial}{\partial x_j}(\bar{u}_i \bar{u}_j) = -\frac{\partial \bar{p}}{\partial x_i} + \frac{\partial \tau_{ij}^{\bar{v}}}{\partial x_j} + \frac{\partial(-\tau_{ij}^r)}{\partial x_j} + \frac{\partial \tau_{ij}^N}{\partial x_j} \tag{2}$$

where $(\bar{\ })$ denotes the filtering operation, $u_i$ the velocity vector, $t$ the time, $p$ the (kinematic) pressure, $\tau_{ij}^{\bar{v}}$ the (kinematic) Newtonian-Like Viscous (NLV) stress tensor, $\tau_{ij}^r$ the deviatoric part of (kinematic) SGS stress tensor, and $\tau_{ij}^N$ the (kinematic) Non-Newtonian SGS (NNSGS) stress tensor. A "kinematic" variable refers to the corresponding variable divided by the fluid density, $\rho$. Hereafter, the word kinematic is dropped and taken implicit for all pressures, viscosities, and stresses. The stresses on the right-hand-side of Eq. (2) are defined by:

$$\tau_{ij}^{\bar{V}} = 2\bar{v}\bar{S}_{ij} \tag{3}$$

$$\tau_{ij}^r = \tau_{ij}^R - \frac{1}{3}\tau_{kk}^R \delta_{ij} \ ; \ \tau_{ij}^R = \overline{u_i u_j} - \bar{u}_i \bar{u}_j \tag{4}$$

$$\tau_{ij}^N = 2\overline{vS_{ij}} - 2\bar{v}\bar{S}_{ij} \tag{5}$$

$$\bar{S}_{ij} = \frac{1}{2}\left(\frac{\partial \bar{u}_i}{\partial x_j} + \frac{\partial \bar{u}_j}{\partial x_i}\right) \tag{6}$$

where $S_{ij}$ is the strain-rate tensor, $\tau_{ij}^R$ the SGS stress tensor, and $\delta_{ij}$ the identity tensor. For a power-law non-Newtonian fluid, the molecular viscosity is given by:

$$v = K_v S^{n-1} \tag{7}$$

$$S = (2S_{ij}S_{ij})^{\frac{1}{2}} \tag{8}$$

where $S$ is the second invariant of the strain-rate tensor, and $n$ and $K_v$ are the power-law index and consistency factor, respectively. For a non-Newtonian fluid, in addition to the conventional SGS stress, $\tau_{ij}^r$, the filtering operation imposes two additional unclosed terms in Eq. (2), i.e., the filtered apparent viscosity, $\bar{v}$, and NNSGS tensor, $\tau_{ij}^N$. For a Newtonian turbulent flow, $\bar{v} = v$ and $\tau_{ij}^N$ vanishes.

Here, the Dynamic Lagrangian (DL) model [32] has been adopted to close the SGS stress as:

$$-\tau_{ij}^r = 2v_r \bar{S}_{ij} \tag{9}$$

where $\bar{S} = (2\bar{S}_{ij}\bar{S}_{ij})^{1/2}$, and $v_r$ is the SGS eddy-viscosity, which is defined by:

$$v_r = C_D \bar{\Delta}^2 \bar{S} \tag{10}$$

where $\bar{\Delta}$ is the primary (or main) filter width and the model coefficient, $C_D$, is estimated dynamically by:

$$C_D = \frac{\varphi_{LM}}{\varphi_{MM}} \quad (11)$$

where $\varphi_{LM}$ and $\varphi_{MM}$ parameters are obtained by the solution of the following transport equations:

$$\frac{\partial \varphi_{LM}}{\partial t} + \bar{u}_j \frac{\partial \varphi_{LM}}{\partial x_j} = \frac{1}{T}(L_{ij}M_{ij} - \varphi_{LM}) \quad (12)$$

$$\frac{\partial \varphi_{MM}}{\partial t} + \bar{u}_j \frac{\partial \varphi_{MM}}{\partial x_j} = \frac{1}{T}(M_{ij}M_{ij} - \varphi_{MM}) \quad (13)$$

$$L_{ij} = \widetilde{\bar{u}_i \bar{u}_j} - \tilde{\bar{u}}_i \tilde{\bar{u}}_j \quad (14)$$

$$M_{ij} = 2\bar{\Delta}^2 \widetilde{|\bar{S}|\bar{S}_{ij}} - 2\tilde{\bar{\Delta}}^2 |\tilde{\bar{S}}|\tilde{\bar{S}}_{ij} \quad (15)$$

$$T = \theta \bar{\Delta}(\varphi_{LM}\varphi_{MM})^{-\frac{1}{8}} \quad (16)$$

In Eqs. (12)-(16), $(\widetilde{\phantom{x}})$ indicates the second filter or test filter. Assuming the primary filter to be the implicit grid filter, the primary filter width is computed by $\bar{\Delta} = \Delta V^{1/3}$ where $\Delta V$ is the local grid cell volume. $\widetilde{\overline{(.)}}$ indicates the double filter operation with the filter width of $\tilde{\bar{\Delta}}$, which is approximated by $\tilde{\bar{\Delta}} \approx \tilde{\Delta} = 2\bar{\Delta}$, and $\theta = 1.5$ is the model constant.

Several previous LES works pointed out that considering $\tau_{ij}^N$ has minor effects on flow predictions [31, 33, 34]. Therefore, similar to the majority of previous TNNF LES studies, $\tau_{ij}^N$ is neglected here, by default. This can be justified when filter width is fine enough everywhere, e.g., in a well-resolved LES, where a large portion of the turbulence kinetic energy is resolved. It should be mentioned that this is not true for RANS, and the counterpart of $\tau_{ij}^N$ has been found to be important in RANS closures [35]. Based on a similar discussion, in the majority of previous studies, the simplifying assumption $\overline{\nu(S)} \approx \nu(\bar{S})$ was considered, which is used here by default:

$$\bar{\nu} \approx K_\nu (\bar{S})^{n-1} \quad (17)$$

An alternative LES approach proposed in our previous work [30], accounting for $\tau_{ij}^N$ and modeling $\bar{v}$ based on a non-linear approximation, is also considered to examine the aforementioned default assumptions. In the alternative approach, the filtered viscosity is approximated by solving the following implicit non-linear equation for $\bar{v}$ using the Newton-Raphson method:

$$\bar{v} - K_v \left[\bar{S}^2 \left(1 + \frac{|v_r|}{\bar{v}}\right)\right]^{\frac{n-1}{2}} = 0 \tag{18}$$

The dynamic Smagorinsky-like model for NNSGS proposed in reference [30] is extended to a dynamic Lagrangian model to relax the limitation of existing homogeneous directions for smoothing the computed dynamic model coefficient. The extended model closure can be summarized as [31]:

$$\tau_{ij}^N = 2v_N \bar{S}_{ij} \tag{19}$$

$$v_N = C_N \bar{\Delta}^2 \bar{S} \tag{20}$$

$$C_N = \frac{\varphi_{LM}^N}{\varphi_{MM}^N} \tag{21}$$

where $\varphi_{MM}^N = \varphi_{MM}$ (given by Eq. (13)), and $\varphi_{LM}^N$ is obtained by solving the following transport equation:

$$\frac{\partial \varphi_{LM}^N}{\partial t} + \bar{u}_j \frac{\partial \varphi_{LM}^N}{\partial x_j} = \frac{1}{T^N}\left(l_{ij}^N M_{ij}^N - \varphi_{LM}^N\right) \tag{22}$$

$$l_{ij}^N = 2\widetilde{\overline{v}\bar{S}_{ij}} - 2\widetilde{\overline{v}}\,\tilde{\bar{S}}_{ij} \tag{23}$$

$$M_{ij}^N = -M_{ij} \tag{24}$$

$$T^N = \theta \bar{\Delta} (\varphi_{LM}^N \varphi_{MM}^N)^{-\frac{1}{8}} \tag{25}$$

## 2.2. The wall models

For TNNF applications, the value of filtered total shear stress $(\tau_{ij}^{\bar{v}} - \tau_{ij}^{r} + \tau_{ij}^{N})$ at a wall, based on the closures described in section 2.1, can be given by:

$$\bar{\tau}_w = \left(\bar{\tau}_{w,21}^2 + \bar{\tau}_{w,23}^2\right)^{1/2} \tag{26}$$

$$\bar{\tau}_{w,2i} = 2\nu_{w,\text{eff}} \bar{S}_{w,2i}; \quad i = 1,3 \tag{27}$$

$$\nu_{\text{eff}} = \bar{\nu} + \nu_r + \nu_N \tag{28}$$

where subscripts 1, 2, and 3 denote the mean stream-wise ($x_1$), wall-normal ($x_2$ or $y$), and span-wise directions ($x_3$), respectively. Since there is a steep velocity gradient near a wall, calculating the wall shear stress based on Eqs. (27) and (28) necessitates a fine computational mesh, which ends up with the approach followed in a WRLES. However, in WSM-WMLES techniques due to the much coarser grid resolution near the walls, the total wall shear stress cannot be computed by Eqs. (27) and (28); on the contrary, the correct value of $\bar{\tau}_w$, which is desired in numerical discretization of the governing equations, e.g., directly in a Finite Volume Method (FVM) or indirectly in a Finite Difference Method (FDM), is imposed through a WSM closure. Using an FVM discretization and assuming the thin boundary layer approximation ($\partial/\partial x_2 \gg \partial/\partial x_1, \partial/\partial x_3$), the prescribed value of the wall shear stress can be imposed by Eq. (27) in conjunction with regulating the value of effective eddy-viscosity at the walls by:

$$\nu_{w,\text{eff}} = \frac{\bar{\tau}_{w,WSM}}{\left[\left(\frac{\partial \bar{u}_1}{\partial x_2}\right)_w^2 + \left(\frac{\partial \bar{u}_3}{\partial x_2}\right)_w^2\right]^{\frac{1}{2}}} \tag{29}$$

where $\bar{\tau}_{w,WSM}$ in Eq. (29) is given by a WSM closure described in the subsequent sections. In the following formulations, the non-dimensional variables include the (normalized) distance from the wall, $y^+ = y u_\tau / \nu_w$, velocity, $u_i^+ = \langle u_i \rangle / u_\tau$, viscosity $\nu^+ = \langle \nu \rangle / \nu_w$, turbulent kinetic energy, $k^+ = k/u_\tau^2$, RMS velocities, $u_i'^+ = \langle u_i'^2 \rangle^{1/2} / u_\tau$, and stresses, $\tau_{ij}^+ = \tau_{ij}/u_\tau^2$; where $\langle \ \rangle$ denotes the ensemble averaging operator, $u_\tau = \tau_w^{1/2}$ the friction velocity, $\tau_w$ the mean wall shear stress, and $\nu_w$ the mean viscosity at the wall.

*2.2.1. Algebraic Werner-Wengle*

In the algebraic Werner-Wengle (WW) law of the wall, it is supposed that the averaged velocity follows the linear-law relation in the region $y^+ \leq 11.81$ and a power-law relation otherwise:

$$u^+ = \begin{cases} y^+ & ; y^+ \leq 11.81 \\ A(y^+)^B & ; y^+ > 11.81 \end{cases} \quad (30)$$

where $u$ refers to the magnitude of the mean velocity relative to the wall, and $A = 8.3$ and $B = 1/7$ are the model constants. It should be mentioned that Eq. (30) is based on averaged velocity ($u^+ = \langle u \rangle / u_\tau$) whose computation is complicated for a general problem with no homogeneous direction. Here, the common practice of using WSM closures with filtered velocity instead of the averaged one, i.e., $u^+ \sim \bar{u}/u_\tau$ in Eq. (30), is incorporated [16]. This principle applies to the other wall models used in this study as well. As an input of Eq. (30), the magnitude of the filtered velocity ($\bar{u}$) is sampled at the height $y = h$ on the wall surface. Afterwards, the WSM equation (Eq. (30)) is solved for $u_\tau = \bar{\tau}_{w,\text{WSM}}^{1/2}$ using the Newton-Raphson root-finding technique. The computed $\bar{\tau}_{w,\text{WSM}}$ is then substituted into Eq. (29) to complete the algorithm. In contrast to Newtonian fluids, for TNNF there is another closure in WSM, namely $\nu_w$ (in $y^+ = yu_\tau/\nu_w$). In a general case, this closure is provided by the power-law fluid nominal wall viscosity relation [36]:

$$\nu_w = K_\nu^{1/n} \tau_w^{1-1/n} \quad (31)$$

and the WSM model should be solved coupled with Eq. (31). In our test cases, i.e., fully-developed pipe flows, $\nu_w$ is known *a priori,* substituting Eq. (48) into (31).

*2.2.2. Integrated Werner-Wengle*

In the framework of collocated FVM discretization, it is argued that the integrated form of WSM relation is more accurate than its algebraic counterpart since the cell-center values represent the volume-integral of fields on a computational cell. By integrating Eq. (30) from the wall ($y = 0$) to the upper edge of the wall-adjacent computational cell ($y = \Delta y_P \sim y_P + 0.5 \times \Delta V_P^{1/3}$ where subscript $P$ indicates the center of the wall-adjacent cell), the integrated form of WW relation is obtained. In the case of the integrated WW WSM, the value of $\bar{\tau}_w$ can be recast explicitly, obviating the need for an iterative root-finding algorithm to solve for $\bar{\tau}_{w,\text{WSM}}$, as:

$$\bar{\tau}_w = \begin{cases} \dfrac{2v_w|\bar{u}_P|}{\Delta y_P} & ; |\bar{u}_P| \leq \dfrac{v_w}{2\Delta y_P} A^{\frac{2}{1-B}} \\ \left[\dfrac{1+B}{A}\left(\dfrac{v_w}{\Delta y_P}\right)^B |\bar{u}_P| + \dfrac{1-B}{2} A^{\frac{1+B}{1-B}} \left(\dfrac{v_w}{\Delta y_P}\right)^{1+B}\right]^{\frac{2}{1+B}} & ; \text{otherwise} \end{cases} \quad (32)$$

Note that integral WSM models are not limited to the use for the first wall-adjacent cell and can be implemented across several cells provided that multiple samples can be extracted from the solution. This requires an elaborate search algorithm in the case of complex unstructured grids which poses a further computational burden. To be comparable with algebraic and ODE models in terms of the number of input samples and computational cost, we limit the present integral WSM models to use only one, wall-adjacent cell, input sample.

*2.2.3. ODE*

By filtering the TBL equation, assuming the thin boundary layer and eddy-viscosity approximations, the filtered TBL equation is derived as:

$$\frac{\partial}{\partial x_2}\left[(\bar{v} + v_r + v_N)\frac{\partial \bar{u}_i}{\partial x_2}\right] = F_i \quad , \quad i = 1,3 \quad (33)$$

where $F_i$ is defined by:

$$F_i = \frac{\partial \bar{p}}{\partial x_i} + \frac{\partial \bar{u}_i}{\partial t} + \frac{\partial \bar{u}_i \bar{u}_j}{\partial x_j} \tag{34}$$

In WMLES, a few off-the-wall cells, covering almost the entire inner part of TBL, are large enough to see a large number of near-wall eddies from a time step to the next, it is sufficient to model the average effect of the interaction of these eddies with the outer layer flow structures, meaning the dynamics of the inner layer can be considered in the mean sense and treated by a RANS framework [37]. Therefore, the approximation $v_r \sim v_t$ in the inner layer is incorporated where $v_t$ is a RANS-like eddy viscosity. For the conventional ODE WSM, $v_N$ is neglected and a linear assumption (Eq. (17)) is used for $\bar{v}$. However, these simplifications are highly questionable for WMLES adjacent to walls due to the large grid (filter) size within this region. Therefore, it is essential to check the sensitivity of results to the modeling of $\bar{v}$ and $v_N$ viscosities. This point is covered in the model described in section 2.2.6. Assuming constant $F_i$ across TBL and integrating Eq. (33) from the wall to point $y = h$ in the wall-normal direction, the two components of the wall shear stress, required to close Eqs. (26) and (29), can be obtained from:

$$\bar{\tau}_{w,2i} = \frac{\bar{u}_i(y=h) - F_i \int_0^h \frac{x_2}{\bar{v}+v_t} dx_2}{\int_0^h \frac{1}{\bar{v}+v_t} dx_2} \tag{35}$$

In the conventional ODE model, developed for Newtonian fluid flows in most CFD packages, in the computation of the integral terms in Eq. (35), $\bar{v}$ is assumed constant and equal to its value on the wall, computed by Eq. (17). The closure for $v_t$ is attained by a mixing length theory with a damping function, which has been proposed by Cabot and Moin [38]:

$$v_t = \bar{v}_w \kappa y^+ \left(1 - \exp\left(-\frac{y^+}{A^+}\right)\right)^2 \tag{36}$$

where $A^+ = 18$ and $\kappa = 0.4$ is the von-Karman constant. Note that Eq. (36) is in fact the extension of Cabot and Moin [38] relation to non-Newtonian fluids where $v$ is replaced with $v_w$ in the definition of non-dimensional quantities, in accordance with the argument made by Edwards and Smith [39]. Eqs. (26), (35), and (36) are coupled and solved by an iterative algorithm. It has been declared that the computational cost of this ODE method is cheaper by over 90 percent compared to the WRLES, which is attributed to fewer computational cells and larger time steps required [40].

*2.2.4. Algebraic Non-Newtonian Shenoy-Saini (NNSS)*

For TNNF in pipes, several experimental correlations have been offered for the mean velocity profile [41-44]. Here, the correlations provided by Dodge and Metzner [43], Clapp [42], and Shenoy and Saini [44] have been examined against the DNS data by Singh, et al. [36]. Among them, we found that the best fit is the one by Shenoy and Saini [44]. They divided the boundary layer into the laminar sub-layer region and the turbulent core. Then, they suggested that velocity in the laminar sub-layer is governed by:

$$u^+ = \hat{y}^{1/n} \; ; \hat{y} \leq \hat{y}_m \tag{37}$$

and in the turbulent core ($\hat{y} > \hat{y}_m$) is given by:

$$u^+ = 2.46 n^{0.25} \left[ \ln(\hat{y})^{\frac{1}{n}} + \left( 0.1944 - \frac{0.1313}{n} + \frac{0.3876}{n^2} - \frac{0.0109}{n^3} \right) \exp\left( \frac{-n^2 \left( \frac{y}{R} - 0.8 \right)^2}{0.129} \right) \right. $$
$$\left. + \frac{1.3676}{n} + \ln 2^{\frac{2+n}{2n}} \right] - \frac{0.4\sqrt{2}}{n^{1.2}} \tag{38}$$

where $R$ is the pipe radius and $\hat{y}$ is the extension of $y^+$ to the power-law fluid as [44, 45]:

$$\hat{y} = \frac{(\tau_w^{2-n})^{\frac{1}{2}}}{K_v} y^n = (y^+)^n \tag{39}$$

where the second equality can be established using Eq. (31). $\hat{y}_m$ is found by the intersection of the laminar sub-layer and turbulent core profiles. For a Newtonian fluid, $\hat{y}$ reduces to $y^+$. Sampling $\bar{u}$ at $y = h$, Eqs. (38) and (39) are solved for $\bar{\tau}_{w,\text{WSM}}$ by an iterative procedure.

### 2.2.5. Integrated Non-Newtonian Shenoy-Saini (NNSS)

By integrating the two-layer *Shenoy-Saini* function over the wall-adjacent computational cell, $\Delta y_P$, the integrated form of *Shenoy-Saini* relation is obtained as:

$$\bar{\tau}_w = K_v \left(\frac{2|\bar{u}_P|}{h}\right)^n \quad ; \hat{y} \leq \hat{y}_m \tag{40}$$

and if $\hat{y} > \hat{y}_m$:

$$|\bar{u}_P| = \frac{y_m^2}{2\Delta y_P}\left(\frac{\bar{\tau}_w}{K_v}\right)^{\frac{1}{n}} + a\left[\left(1 - \frac{y_m}{\Delta y_P}\right)\ln\left(\left(\frac{\bar{\tau}_w^{2-n}}{K_v^2}\right)^{\frac{1}{2n}}\right) + \ln \Delta y_P + \frac{y_m}{\Delta y_P}[1 - \ln y_m] - 1\right]\bar{\tau}_w^{1/2}$$

$$+ \frac{Rb}{2\Delta y_P}\left(\frac{\pi}{c}\right)^{\frac{1}{2}}\left[\text{erf}\left[\sqrt{c}\left(\frac{\Delta y_P}{R} - 0.8\right)\right] - \text{erf}\left[\sqrt{c}\left(\frac{y_m}{R} - 0.8\right)\right]\right]\bar{\tau}_w^{1/2} \tag{41}$$

$$+ d\left(1 - \frac{y_m}{\Delta y_P}\right)\bar{\tau}_w^{1/2}$$

where the constants $a$, $b$, $c$, and $d$ are as follows:

$$\begin{aligned}
a &= 2.46 n^{0.25} \\
b &= a\left[0.1944 - \frac{0.1313}{n} + \frac{0.3876}{n^2} - \frac{0.0109}{n^3}\right] \\
c &= \frac{n^2}{0.129} \\
d &= a\left[\frac{1.3676}{n} + \ln 2^{\frac{2+n}{2n}}\right] - \frac{0.4\sqrt{2}}{n^{1.2}}
\end{aligned} \tag{42}$$

Eq. (41) is implicit for $\bar{\tau}_{w,\text{WSM}}$ and solved by an iterative procedure.

### 2.2.6. Non-Newtonian ODE (NNODE)

In this section, an improved variant of the well-known standard Newtonian ODE WSM, described in section 2.2.3, is proposed by relaxing the linear model assumption for $\bar{v}$ and accounting for $\nu_N$. Assuming constant $F_i$ across TBL and integrating Eq. (33) from the wall to point $y = h$ in the wall-normal direction, the two components of the wall shear stress in Eq. (26) can be obtained from:

$$\bar{\tau}_{w,2i} = \frac{\bar{u}_i(y=h) - F_i \int_0^h \frac{x_2}{\bar{v}+v_t+v_N} dx_2}{\int_0^h \frac{1}{\bar{v}+v_t+v_N} dx_2} \tag{43}$$

A choice for modeling $\bar{v}$ in Eq. (43) is to use the same closures proposed for the main LES away from the wall, i.e., the linear (Eq. (17)) or non-linear (Eq. (18)) models. The linear model is used in the conventional ODE model (section 2.2.3). However, these closures are based on the value of the mean shear rate, $\bar{S}$, which is not well-resolved near the walls in WMLES due to the coarse grid size in this region. Another approach, which is followed here in the NNODE model, is assuming $\bar{v}$ in Eq. (43) to be constant and approximated by the local mean value on the wall ($\bar{v} \approx v_w$) given by Eq. (31). This assumption can be justified considering the DNS results by Singh, et al. [36] showing the relatively moderate change of $\bar{v}$ near the walls. The advantage of this approximation over the linear or non-linear models is that it is not based on the value of $\bar{S}$ which carries large errors near the walls in a WMLES.

The turbulent eddy viscosity, $v_t$, in Eq. (43) is also modeled by Eq. (36) where $\bar{v}_w$ in this equation is again computed here by Eq. (31). As discussed earlier, the omission of $v_N$ adopted in the standard ODE model, is questionable. To relax this assumption in the present NNODE model, we use a procedure similar to the one proposed by Gavrilov and Rudyak [35] in the context of 4-equation RANS models as:

$$v_N = (n-1)\frac{\varepsilon}{\langle S^2 \rangle}, \quad \langle S^2 \rangle = \langle S \rangle^2 + \frac{\varepsilon}{v_w} \tag{44}$$

where $\varepsilon$ is the turbulent dissipation rate. Here, the equilibrium assumption is used in the inner layer, and the dissipation rate can be estimated by:

$$\varepsilon \sim \text{Pr} = \left| -\tau_{ij}^t \langle S_{ij} \rangle \right| = \left| -2v_t \langle S_{ij} \rangle \langle S_{ij} \rangle \right| = v_t \langle S \rangle^2 \tag{45}$$

where Pr means the rate of production of the turbulent kinetic energy. Substituting Eq. (45) into Eq. (44), the following closure is obtained for $v_N$:

$$\nu_N = (n-1)\frac{\nu_t}{1+\frac{\nu_t}{\bar{\nu}_w}} \tag{46}$$

Eqs. (26), (36), (31), (43), and (46) are coupled and solved by an iterative algorithm.

## *2.3. Description of test cases*

There are several turbulent non-Newtonian pipe flow DNS databases in the literature [34, 36, 45-47]. The database by Singh, et al. [36] is considered here as a reference due to its relatively higher Reynolds numbers and the details provided on each case.

The computational geometry is a pipe with the length $L_z = 8\pi R$ and radius $R = 0.5\ m$. The results are analyzed in the cylindrical $(z, r, \theta)$ coordinates. The governing non-dimensional parameters are the power-law index, $n$, and friction Reynolds number, $Re_\tau$, which is defined by:

$$Re_\tau = \frac{u_\tau R}{\nu_w} \tag{47}$$

To assess the performance of different models, two cases from this database are chosen here, one Newtonian case ($n = 1$) and one shear-thinning case ($n = 0.6$). The parameters of these test cases are reported in table 1. For the simulation of fully-developed pipe flow, the mean pressure gradient along the pipe, $dP/dz$, is applied as an extra source term on the right-hand side of Eq. (2). $Re_G$, $Re_{MR}$, and $U_b$ are the generalized Reynolds number, Metzner-Reed Reynolds number, and the pipe bulk velocity, respectively[36].

Table 1: The information of the DNS reference cases.

| Governing parameters | | dimensional inputs | | Other useful information [36] | | |
|---|---|---|---|---|---|---|
| $n$ | $Re_\tau$ | $K_\nu$ $(m^2/s^{2-n})$ | $dP/dz$ $(m/s^2)$ | $Re_G$ | $Re_{MR}$ | $U_b$ $(m/s)$ |
| 0.6 | 323 | 0.05429 | 16 | 11093 | 5498 | 34.46 |
| 1.0 | 323 | 0.00309 | 16 | 10322 | 10322 | 31.86 |

Note that for the present fully-developed pipe flow test cases, the mean wall apparent viscosity and shear stress are constant and can be computed *a priori* via Eq. (31) and the relation [36]:

$$\tau_w = \frac{R}{2}\frac{dP}{dz} \tag{48}$$

Therefore, instead of $\tau_w$, the pipe bulk velocity, $U_b$, prediction can be regarded as a global measure of the accuracy of WSM closures.

The flow statistics are reported based on the ensemble (Reynolds) averaging in DNS databases, e.g., the mean velocity, $\langle u_i \rangle$, Reynolds stress, $\langle u_i' u_j' \rangle$ where $u_i' = u_i - \langle u_i \rangle$, etc. These quantities should be computed from filtered fields in a LES. For this purpose, the assumption $\langle q \rangle \approx \langle \bar{q} \rangle$ is used [48] for the estimation of Reynolds-averaged quantities. By this assumption, it can be shown that the mean velocity and Reynolds stress can be calculated from LES data by:

$$\langle u_i \rangle \approx \langle \bar{u}_i \rangle; \quad \langle v \rangle \approx \langle \bar{v} \rangle; \quad \langle S_{ij} \rangle \approx \langle \bar{S}_{ij} \rangle \tag{49}$$

$$\langle u_i' u_j' \rangle \approx \langle (\bar{u}_i - \langle \bar{u}_i \rangle)(\bar{u}_j - \langle \bar{u}_j \rangle) \rangle + \langle \tau_{ij}^r \rangle + \frac{2}{3}\langle k_{\text{SGS}} \rangle \delta_{ij} \tag{50}$$

where $k_{\text{SGS}}$ is the SGS kinetic energy. The first term on the right-hand side of Eq. (50) constitutes the resolved part and the next two terms compose the unresolved part which requires modeling. The closure of $\tau_{ij}^r$ is based on the chosen SGS closure (see section 2.1) and $k_{\text{SGS}}$ is approximated, for the DL LES model, by [49]:

$$k_{\text{SGS}} \approx \left(\frac{C_{\text{D}}}{C_{\text{e}}}\right)^{\frac{2}{3}} \bar{\Delta}^2 \bar{S}^2 = \left(\frac{\nu_r}{C_{\text{e}}^{1/3} C_{\text{D}}^{2/3} \bar{\Delta}}\right)^2 \tag{51}$$

where the model constant is $C_{\text{e}} = 1.048$.

## 3. Numerical methods

The WRLES and WMLES simulations have been performed based on the cell-centered collocated FVM in OpenFOAM open-source CFD package (www.openfoam.org), version 4.x. The Pressure-Implicit with Splitting of Operators (PISO) algorithm [50] was used for pressure-velocity coupling with 2 pressure correction loops. The implicit second-order "backward" Euler scheme [51, 52] was employed to discretize time derivatives. The Green-Gauss cell-based "Gauss linear" scheme [51] was adopted for gradients. The momentum advection term was discretized by the "Gauss linear" centered scheme of the second-order accuracy. $\varphi_{LM}$ and $\varphi_{MM}$ advection terms were discretized by "Gauss limitedLinear 1" scheme, which is the bounded central difference based on Total Variation Diminishing (TVD) [53] that avoids overshoot/undershoot problems. The "Gauss linear corrected" scheme, in which non-orthogonality effects (or cross-diffusion) are explicitly treated based on the over-relaxed approach [54], was incorporated to discretize the diffusion terms. Lastly, the interpolated variables on the cell faces are obtained by the "linear" scheme. To solve the system of discretized linear algebraic equations, the Geometric Agglomerated algebraic MultiGrid solver (GAMG) along with Diagonal Incomplete-Cholesky with Gauss-Seidel (DICGaussSeidel) preconditioner [55] was applied for the pressure. For the momentum, $\varphi_{LM}$, and $\varphi_{MM}$ transport equations, the Stabilized Preconditioned Bi-Conjugate Gradient (PBiCGStab) solver [56, 57] with Diagonal Incomplete-LU (DILU) preconditioning [55] was utilized. A normalized residual tolerance of $10^{-6}$ is set as the convergence criterion for all variables at each time step. The WMLES algorithms were developed by extending the library prepared and published by Mukha, et al. [16] to include the newly proposed non-Newtonian models. For the numerical computation of integrals in ODE and NNODE models, the trapezoidal rule is utilized by the division of the distance between the wall and center of the sampling cell ($h$) into 30 equal segments. For the test cases considered in this study, NEE can be ignored and an equilibrium assumption of $F_i = 0$ is adopted in ODE and NNODE models.

The inlet and outlet boundary conditions for all the quantities are periodic (cyclic). At walls, the velocity and $\varphi_{LM}$ are zero while $\varphi_{MM}$ and pressure have a zero-gradient. The initial condition for the velocity is set based on the method proposed by De Villiers [58], in which interactions of a particular oscillating velocity in the azimuthal direction and a particular mean axial velocity can trigger turbulence. The initial conditions for quantities $\varphi_{LM}$ and $\varphi_{MM}$ in Eqs. (12) and (13) are constructed according to [32]:

$$\varphi_{MM} = M_{ij}M_{ij} \tag{52}$$

$$\varphi_{LM} = 0.0256\varphi_{MM} \tag{53}$$

The simulation strategy is illustrated in figure 1, where the regulated Courant–Friedrichs–Lewy (CFL) number is plotted against the simulation time. For accelerating the simulation, the Newtonian WRLES starts with the Wall-Adapting Local Eddy-viscosity (WALE) [59] SGS model since it is more robust and faster than DL model. Simulation commences with the low CFL value of 0.1 for a few (50) time steps, then the CFL number is gradually increased to 16 by doubling CFL every 20 time steps. After elapsing the time $T_2 = 14\ FTT$, where $FTT$ is the Flow-Through-Time ($FTT = L/U_b$), the SGS model is switched to DL while CFL is dropped to 0.5, and simulation continues till $T_3 = 18\ FTT$ at which the statistically stationary state prevails. After that, the time averaging process begins and continues for $T_4 - T_3 = 11\ FTT$ to collect the statistics of the flow. The non-Newtonian WRLES is initialized with the results of the previous simulation and continues for $T_{1NN} - T_{0NN} = 3\ FTT$ to reach the stationary state. Additional $T_{2NN} - T_{1NN} = 9\ FTT$ time is required to collect statistics by time averaging. WMLES is initialized with its corresponding WRLES and requires about $3\ FTT$ time for reaching the stationary state, although the required averaging time varies for each WMLES simulation, as mentioned in table 3.

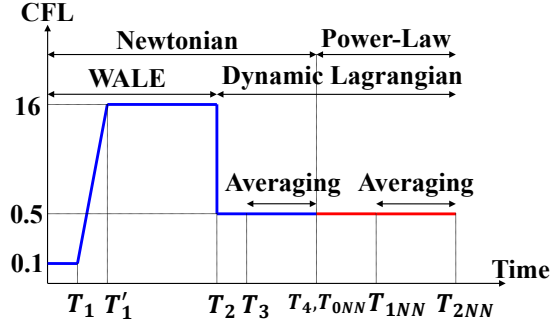

Figure 1: The simulation strategy for WRLES. The CFL setting versus simulation time.

Here, the required period of time-averaging for computing flow statistics is considerably reduced by performing spatial averaging in addition to the time-averaging of the instantaneous data. For this purpose, the spatial averaging on the collected time-averaged data is done in both azimuthal and axial homogeneous directions. For each radial location, spatial samples are taken on all points near the grid cell centers in both axial and azimuthal directions (128×168 samples).

The choice of WRLES computational grid, even for a simple fully-developed turbulent pipe flow, is a delicate task. Some recommendations from other studies have been collected in table 2.

Table 2: Different mesh design criteria for LES and DNS of Newtonian/non-Newtonian pipe flow.

| Reference | Description | Type of simulation | $y_w^+$ | $(r\Delta\theta)_w^+$ | $\Delta z^+$ |
|---|---|---|---|---|---|
| Rudman and Blackburn [46] | Power-law fluid ($n = 0.6$) | DNS | 0.25 | 7 | 11 |
| Gnambode, et al. [33] | Power-law fluid ($n = 0.69$) | WRLES | 0.059 | 12.21 | 38.88 |
| Chin, et al. [60] | Newtonian | DNS | 0.03 | 6.56 | 7.87 |
| | Newtonian | WRLES | 0.048 | 19.6 | 32.8 |
| Kim, et al. [61] | Newtonian | DNS | 0.05 | 7 | 12 |
| Viazzo, et al. [62] | Newtonian | WRLES | 0.88 | 15.71 | 31.4 |

According to the data in table 2, it can be deduced that the cell sizes in the stream-wise and azimuthal directions in LES are, respectively, 3-5 and 2-3 times the ones in DNS. Based on the mesh resolution chosen in the reference DNS solution [36], i.e., $\Delta z^+ = 21$ and $(r\Delta\theta)_w^+ = 4.5$, The grid resolution for WRLES in the present study is chosen as $\Delta z^+ = 63.4 \cong 3(\Delta z^+)_{\text{DNS}}$ and $(r\Delta\theta)_w^+ = 12.08 \cong 3(r\Delta\theta)_{w,\text{DNS}}^+$. A block-structured O-type grid with hexagonal cells is generated for WRLES. The cross-section of the computational grid is shown in figure 2. For the wall-normal direction, $y_w^+ = 0.45$ and the grid expansion factor is between 1 and 1.4. The grid non-orthogonality is about 5-10 percent, which conveys a high-quality mesh. Four cells have been laid in the laminar sub-layer ($y^+ < 5$), hence, the criterion of existing at least three cells in the laminar sub-layer, proposed by Eggels, et al. [63], is satisfied. The number of cells is 128 and 168 in the axial and azimuthal directions, respectively. The number of cells in each pipe cross-section is 9324 and the total number of cells is 1 193 472.

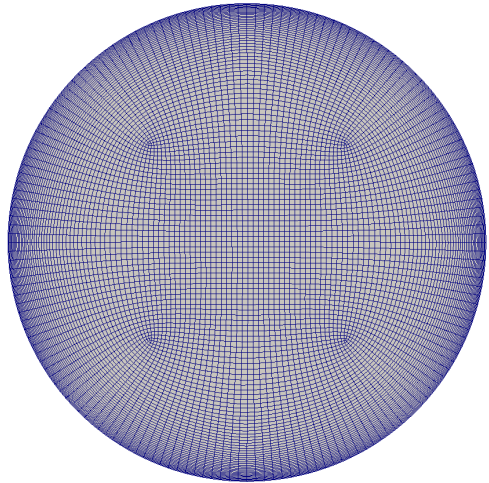

Figure 2: The cross-section of the computational grid for WRLES.

For the validation of our computations, the results of the present WRLES are compared with the reference DNS database in figure 3. To check the consistency of the modeling and grid sensitivity test, the WRLES solution on a finer grid with $\Delta z^+ \sim 34$ and a total number of cells of 2 219 112, is also reported in this figure. As can be seen, by refining the grid (decreasing the filter width of LES) and approaching the DNS limit, the WRLES solution approaches the reference DNS solution [36]. The mean velocity profiles of WRLES on both grids almost coincide with a negligible difference near the center line while there is about a 6% improvement in the prediction of the peak kinetic energy by the fine grid. It should be noted that the filter width of the fine grid WRLES is too small to be considered a reference LES ($(\Delta z^+)_{LES} \cong 1.5(\Delta z^+)_{DNS}$). In other words, the fine grid WRLES solution is a DNS-like solution and the effect of SGS modeling would be minor. Therefore, it cannot be regarded as a challenging test case. Therefore, the medium grid solution described previously, which is also in accordance with the recommendations reported in table 2, is used as our reference WRLES in the rest of this paper. To further support that the medium grid LES can be accepted as a WRLES, its quality criteria have been checked in Appendix A.

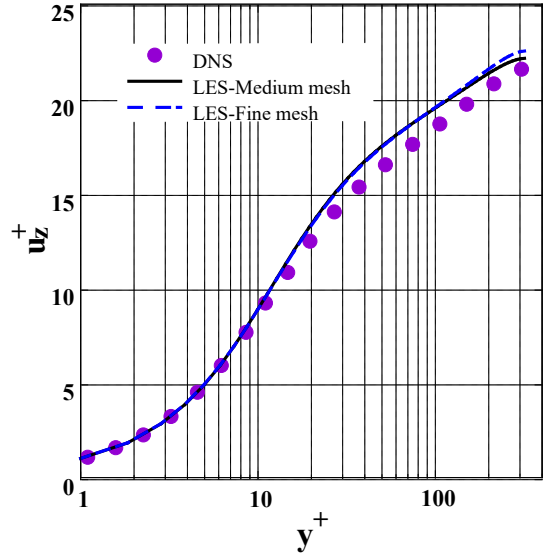

(a)

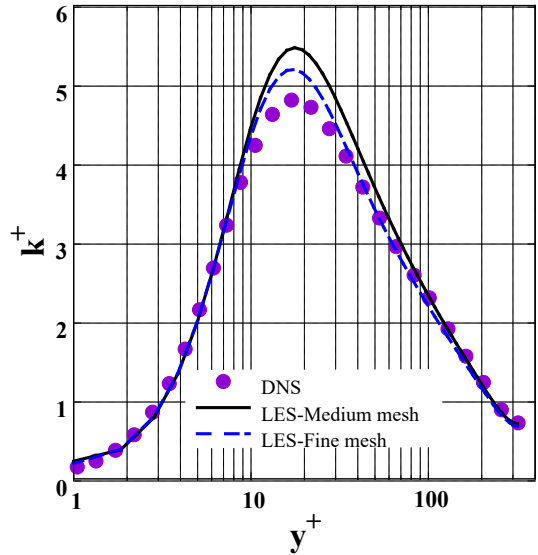

(b)

Figure 3: The shear-thinning case results on the fine and medium (main) grid. The mean axial velocity (a) and total turbulent kinetic energy (b) versus $y^+$. The comparison of current WRLES and reference DNS [36].

## 4. Results and discussion

*4.1. WRLES*

The WRLES solutions are used as the reference to assess the accuracy of WMLES models. To check the validity of different flow statistics predicted by WRLES, these predictions are compared against the DNS data in figure 4 to figure 6. WRLESs have been carried out with different choices of filtered viscosity and NNSGS modeling to assess the effect of modeling strategies as well. As can be observed in figure 4, the first-order flow statistics, i.e., the mean velocity and apparent viscosity, predicted by WRLES are in fine agreement with the DNS data. The shear-thinning effect, the reduction of fluid viscosity with the increase in the shear rate by approaching the pipe walls, has been captured well by WRLES solutions. In addition, it can be deduced that the influence of considering the more complicated non-linear filtered viscosity and NNSGS dynamic closures is negligible for the present WRLES. This is probably due to the choice of the LES filter width which is chosen close to the DNS grid size based on the analysis performed in section 3. This results in the resolution of a large portion of the kinetic energy, leaving only a small part for the SGS modeling. Note that we also tried coarser grids for WRLES to achieve higher LES-to-DNS grid-size ratios or equivalently LES filter widths and assess the performance of non-linear models in those conditions. However, the level of LLM error was so high for the coarse grid LES and completely hid the effect of SGS closures. This highlights the importance of the analysis performed in section 3 to choose the proper grid resolution. It should be noted that, in contrast to WRLES, the negligible effect of non-linear filtered viscosity may not be valid for WMLES due to the large LES-to-DNS grid-size ratio near the walls and improved closures may be necessary in that case. This point will be scrutinized in section 4.4.4.

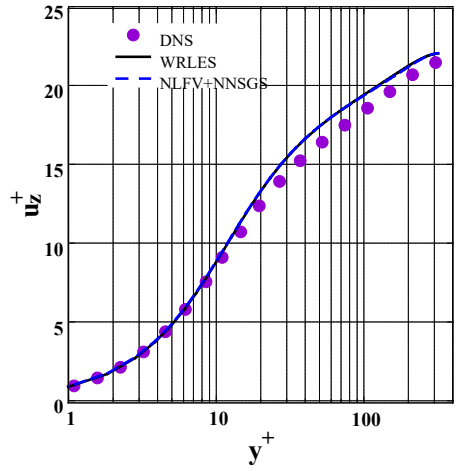

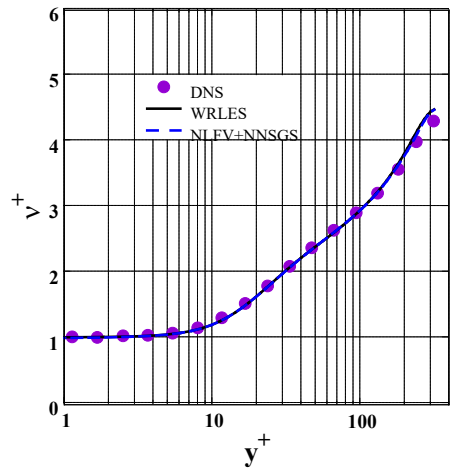

Figure 4: The mean axial velocity (a) and apparent viscosity (b) profiles of the shear-thinning case ($n = 0.6$) obtained by WRLES against DNS data [36]. The effect of filtered viscosity and NNSGS modeling: The default WRLES with the linear filtered viscosity model and omitting NNSGS, and NLFV+NNSGS with the inclusion of non-linear filtered viscosity and dynamic NNSGS models.

In figure 5, the RMS velocity components profiles computed from WRLES are compared against the DNS ones. The non-linear filtered viscosity closure and including the NNSGS model have still an unnoticeable influence on the solutions. The axial RMS component is in satisfactory agreement with the DNS. Nevertheless, there is about a 10% overprediction of its peak value by WRLES. The WRLES radial and azimuthal RMS components show higher degrees of deviation from the DNS ones. However, the axial component is dominant, and therefore, the WRLES turbulent kinetic energy is captured well by WRLES. As can be seen in figure 6, the other second-order statistics, including the shear Reynolds stress and NL stress, are in fine agreement with the DNS predictions.

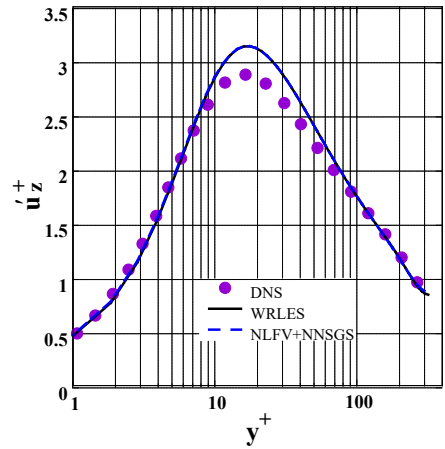

(a)

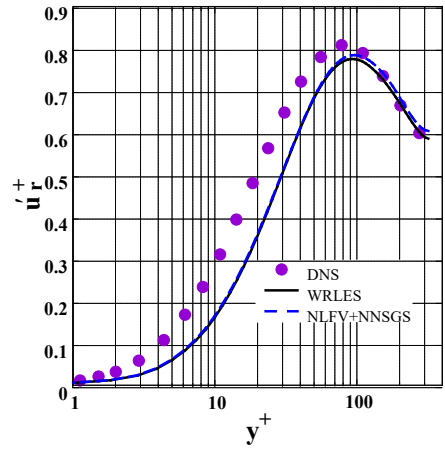

(b)

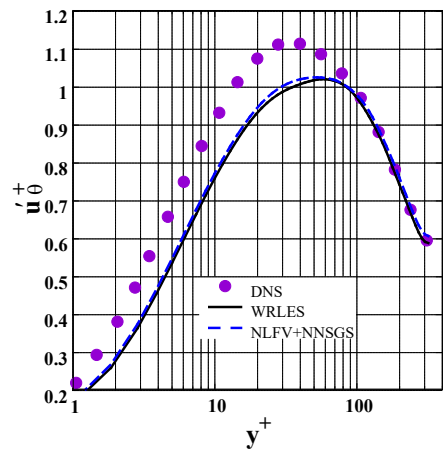

(c)

Figure 5: The profiles of the resolved RMS velocities, $u_i'^+$, in the axial (a), radial (b), and azimuthal (c) directions. The effect of filtered viscosity and NNSGS modeling. The legend is the same as figure 4.

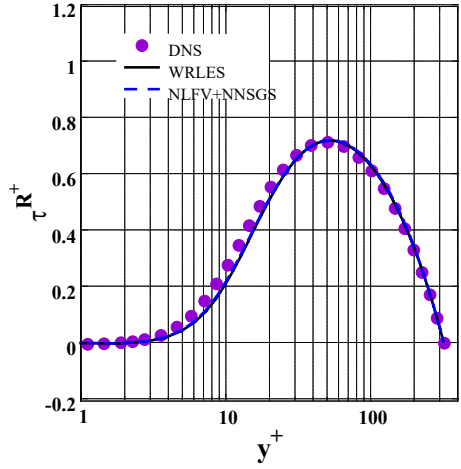

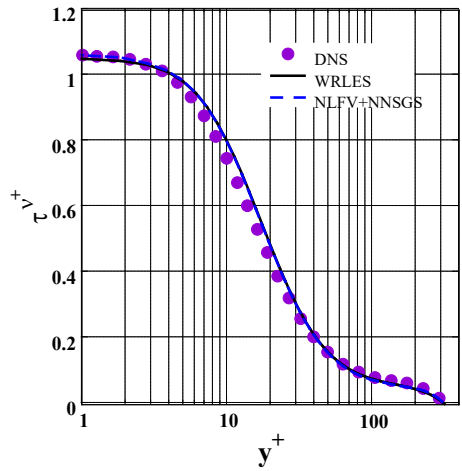

Figure 6: The profiles of the total shear Reynolds stress, $\tau^{R+} = \langle u'_r u'_z \rangle / u_\tau^2$ (a) and mean shear viscous stress, $\tau^{\nu+} = \nu^+ \partial u_z^+ / \partial y^+$ (b). The effect of filtered viscosity and NNSGS modeling. The legend is the same as figure 4.

## *4.2. WMLES case description*

To assess the performance of different WSM model types, parameters, and the effect of grid resolution on WMLES, 22 cases are designed. The specification of these WMLES simulations has been reported in table 3. WMLES simulations are numbered 1-5 for the Newtonian test case and 6-22 for the shear-thinning non-Newtonian case introduced in table 1. In cases 1-5, the effect of the WSM model type (algebraic versus integrated WW), and the height of the sampling point at different regions of the inner layer, including the upper edge of the buffer layer ($h^+ \sim 30$) and within the log-law layer near the lower edge of the overlap region ($h^+ \sim 60$), are examined. Integrated and algebraic WW models are sampled at the first and third cell off-the-wall, respectively, which have been recommended in previous studies, see e.g., reference [16]. Case 5 setting is the same as case 1, with the same WMLES grid while the no-slip condition was applied instead of a WSM model to check the effect of using WMLES.

Table 3: The specification of WMLES simulations. The cross-section of the computational domain for each case is illustrated in figure 7. The CPU time for 1 FTT Newtonian and non-Newtonian WRLES is 24 058 and 25 178 s, respectively.

| Case | Fluid | WSM Model | Height of the sampling point, $h^+$ | Sampling at the $n_s^{th}$ cell away from wall | CPU time for 1 FTT simulation (seconds) | Time averaging period (FTT) |
|---|---|---|---|---|---|---|
| 1 | Newtonian | Algebraic WW | 58 | 3 | 16 298 | 11 |
| 2 | | Algebraic WW | 28.75 | 3 | 18 902 | 11 |
| 3 | | Integrated WW | 60‡ | 1 | 12 246 | 17 |
| 4 | | Integrated WW | 30‡ | 1 | 19 782 | 11 |
| 5 | | No WSM (no-slip) | 58 | 3 | 16 830 | 33† |
| 6 | non-Newtonian, power-law, $n = 0.6$ | Algebraic WW | 58 | 3 | 16 166 | 12 |
| 7 | | Algebraic WW | 28.75 | 3 | 18 960 | 12 |
| 8 | | Integrated WW | 60‡ | 1 | 12 596 | 27 |
| 9 | | Integrated WW | 30‡ | 1 | 17 534 | 21 |
| 10 | | ODE | 58 | 3 | 17 808 | 12 |
| 11 | | ODE | 28.75 | 3 | 22 096 | 9 |
| 12 | | Algebraic NNSS | 58 | 3 | 17 496 | 12 |
| 13 | | Algebraic NNSS | 28.75 | 3 | 19 306 | 15 |
| 14 | | Algebraic NNSS * | 28.75 | 3 | 9188 | 15 |
| 15 | | Algebraic NNSS | 28.75 | 2 | 19920 | 12 |
| 16 | | Algebraic NNSS | 30‡ | 1 | 30 044 | 18 |
| 17 | | Algebraic NNSS | 14.15 | 3 | 22 426 | 18 |
| 18 | | Integrated NNSS | 60‡ | 1 | 15 096 | 18 |
| 19 | | Integrated NNSS | 30‡ | 1 | 19 768 | 15 |
| 20 | | NNODE | 58 | 3 | 14 514 | 15 |
| 21 | | NNODE | 28.75 | 3 | 18 968 | 12 |
| 22 | | No WSM (no-slip) | 58 | 3 | 16 270 | 24† |

\* The grid resolution in the axial direction is coarsened ($\Delta z^+ = 128$), compared to the base case ($\Delta z^+ = 64$).
† The complete convergence in time averaged statistics has not been achieved even with a long time-averaging process.
‡ The upper edge of the sampling cell.

Cases 6-22 are designed to assess the performance of WMLES for non-Newtonian power-law fluids. Cases 6-11 use a standard WSM, developed for Newtonian fluids, for the non-Newtonian case which is the conventional procedure due to the availability of standard models in CFD packages. Case 6 is the base case with the algebraic WW and sampling at the 3$^{rd}$ cell near the lower edge of the overlap region ($h^+\sim 60$). Case 7 checks the effect of lowering the sampling height to the upper edge of the buffer layer ($h^+\sim 30$). In cases 8 and 9, the results of using integrated WW with different $h^+$ values are compared with the previous cases. In cases 10 and 11, the standard ODE model is exploited to check the merit of the ODE model over algebraic/integrated models.

Cases 12 to 21 are designated to evaluate the performance of the present extended non-Newtonian WSM models. Case 12 uses the algebraic NNSS with sampling at the 3$^{rd}$ cell near the lower edge of the overlap region ($h^+\sim 60$). Cases 12, 13, and 17 investigate the effect of changing the sample height from $h^+\sim 60$ to 15; in different areas of the inner layer including the lower edge of the overlap region, the upper edge of the buffer layer, and within the buffer layer. Case 14 is considered to study the effect of grid resolution in the axial direction by doubling the grid size (or equivalently filter width). This case is the same as case 13 except for its $\Delta z^+$ which is considered to be 128 instead of 64. Cases 16, 15, and 13 study the effect of sampling cell from the 1$^{st}$, 2$^{nd}$, and 3$^{rd}$ cell off-the-wall. In cases 18 and 19, the influence of using integrated NNSS with different $h^+$ values is explored, and in cases 20 and 21, the NNODE performance with different $h^+$ values is evaluated. Finally, in case 22, the no-slip condition is exploited instead of WMLES on the same computational mesh as case 6 to delineate the necessity of using WSM for non-Newtonian fluids when WMLES (coarse) grids are used for LES.

One-quarter of the cross-section of the computational grid used for each case of table 3 is illustrated in figure 7, where red cells denote the cells in which the sampling point is located. It should be noted that the grids are designated in such a way that the grid in the outer layer beyond the sampling point is the same as the WRLES grid to minimize the deviation of the WMLES and WRLES results due to the domain discretization away from the wall. It means, only the portion of the grid between the sampling location and the wall is considered to have a much coarser mesh, called a WMLES-type grid. In this near-wall region, the number of cells and their configuration are chosen so as to have the desired $n_s$ and $h^+$ based on table 3. In figure 7d, f, and c, the first, second, and third cells away from the wall have been used as the sampling cell, respectively, and the upper bound of the sampling cells has been set at $y^+\sim30$. In figure 7a, c, and e, the third cell has been exploited as the sampling cell while the upper bound of the sampling cells has been located within the log layer at $y^+\sim60$ (slightly above the lower edge of the overlap region), $y^+\sim30$ (the beginning of the logarithmic layer), and $y^+\sim15$ (within the buffer layer), respectively. In figure 7a and b, the sampling at the third and first cells have been used at the height $y^+\sim 60$.

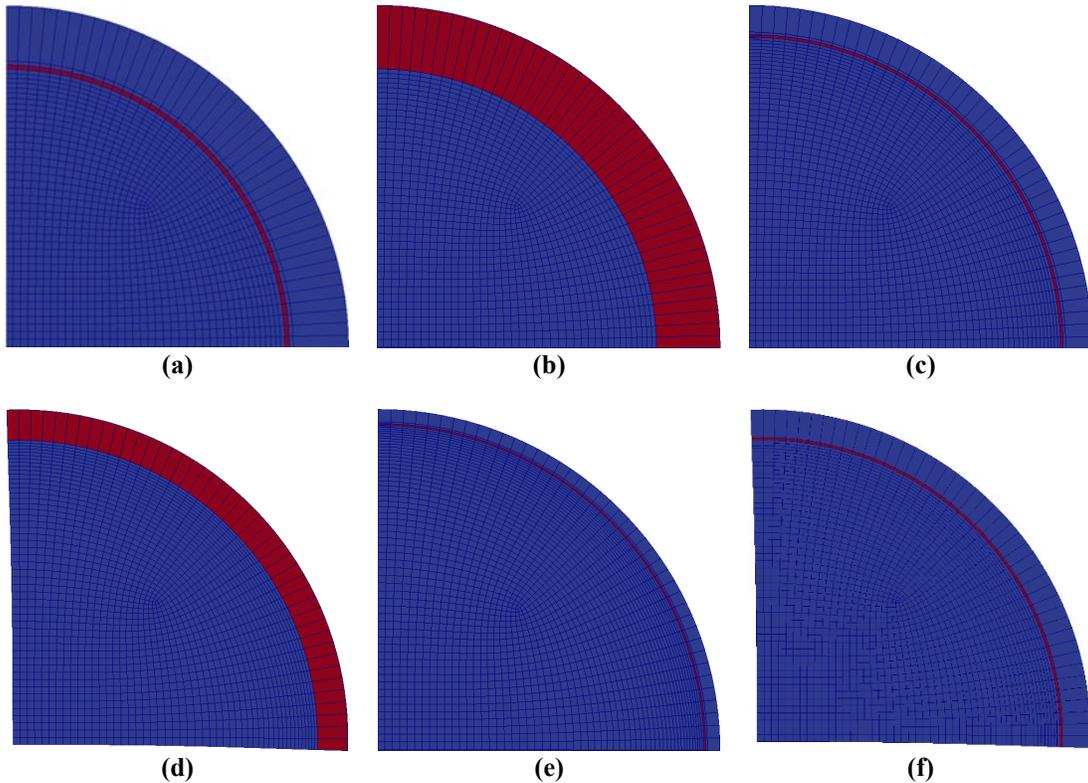

Figure 7 One-quarter of the cross-section of the computational grid used for each WMLES based on table 3. (a) cases 1, 5, 6, 10, 12, 20, and 22, (b) cases 3, 8, and 18, (c) cases 2, 7, 11, 13, 14, and 21, (d) cases 4, 9, 16, and 19, (e) case 17, and (f) case 15. The sampling point is located in the cells colored with red.

### 4.3. WMLES of Newtonian fluid flow

Some comments have to be made on the selected simulation cases reported in table 3. There is strong evidence that algebraic models are ineffective if their sampling is done from the wall adjacent cell, i.e., $n_s = 1$ [16, 24]. Therefore, we do not further examine this point for the Newtonian case, and for the algebraic models, $n_s = 3$ is chosen. In addition, based on many previous investigations, e.g., see reference [23], it is trivial that integral models outperform their corresponding algebraic models when the grid size in the wall-normal direction increases, and this fact is not investigated again. As far as our designed grid is very fine at $n_s = 3$, conclusions for an algebraic model with $n_s = 3$ can be generalized to its corresponding integral method with a single sampling point at $n_s = 3$.

To assess the performance of WSM closures for the Newtonian test case, the results of WMLES cases are compared against the result of WRLES. The predicted mean axial velocity and Reynolds stress components profiles are illustrated in figure 8 and figure 9, respectively. As it can be observed, when the grid is too coarse to capture the dynamics of the inner TBL, the use of WSM is necessary; otherwise, the simple no-slip condition (WMLES5) leads to a large deviation from WRLES even in the primary statistics, namely the mean axial velocity. These large errors have also been reported in previous studies [25]. According to our results, large deviations in turbulence intensities, especially in the stream-wise component, are also observed without implementing a WSM.

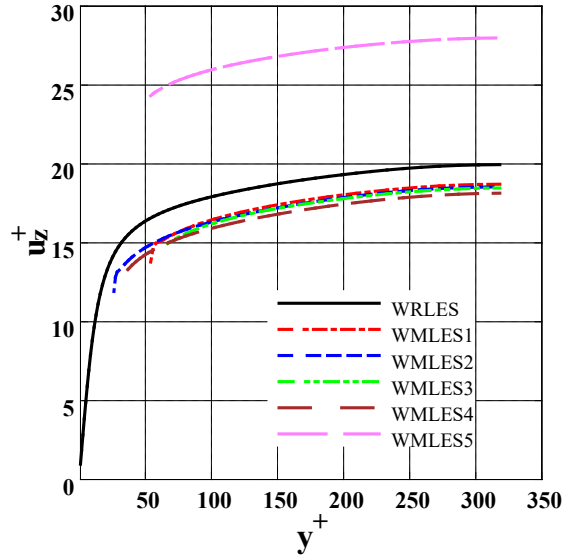

Figure 8: The mean axial velocity profile: The comparison of WMLES cases (see table 3) and WRLES predictions for the Newtonian test case.

Mukha, et al. [16] examined the effect of sample height ($h$) variation within the overlap region for the algebraic Spalding's model [64] and observed no significant sensitivity of mean velocity, turbulent kinetic energy, and shear Reynolds stress to the change of $h$ in the overlap region while keeping $n_s$ constant. Nevertheless, some sensitivity was reported when $h$ increases to the values close to the upper bound of the overlap region. Similarly, based on our results, for both algebraic and integral models, by altering the sampling location from within the log layer near the lower edge of the overlap region ($h^+\sim 60$ in WMLES1 and 3) to the lower edge of the log layer ($h^+\sim 30$ in WMLES2 and 4) the predicted mean velocity, shear Reynolds stress, and axial RMS velocity show no significant sensitivity; however, the radial and azimuthal RMS components predictions improve. Based on the results reported for the shear Reynolds stress, $\tau^{R+}$, in figure 9, using no WSM leads to the under-prediction of $\tau^{R+}$ while implementing a WSM regardless of its type or sampling height, $h$, brings about an accurate prediction of this parameter. This conclusion was also supported by the results obtained in previous studies [16, 23]. The abrupt change at the beginning of Reynolds stress profiles predicted by WMLES models has also been observed elsewhere [25] which is attributed to the uncertainty of the WMLES results in the inner layer, very close to the walls.

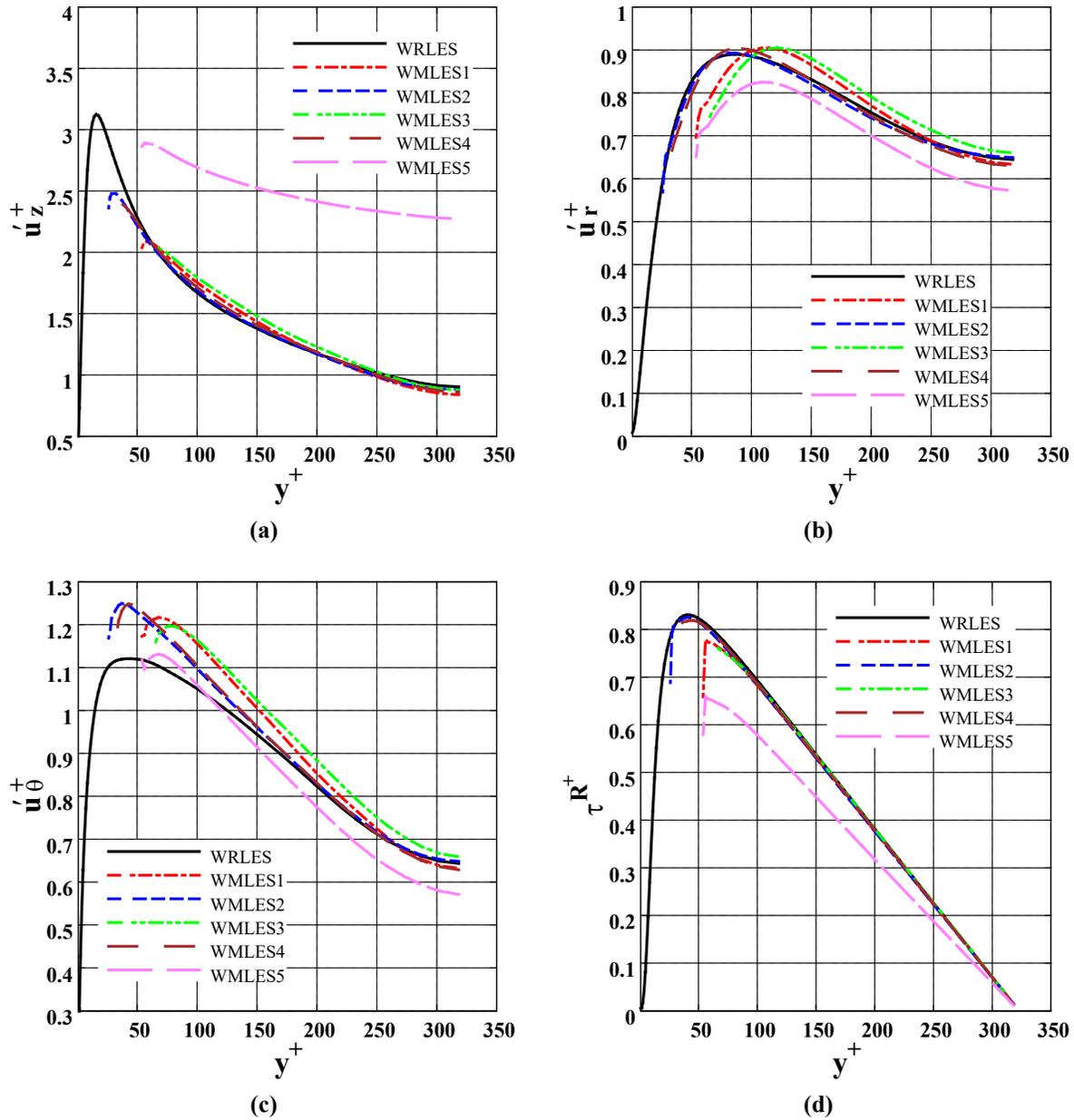

Figure 9: The profile of the resolved RMS velocity, $u_i'^+$, in the axial (a), radial (b), and azimuthal (c) directions, and the total shear Reynolds stress, $\tau^{R+} = \langle u_r' u_z' \rangle / u_\tau^2$, profile (d): The comparison of WMLES cases (see table 3) and WRLES predictions for the Newtonian test case.

Some other interesting conclusions can be drawn from the results reported in figure 8 and figure 9. Although avoiding the sampling from the wall adjacent cell ($n_s = 1$) can improve the results of algebraic models considerably, the sensitivity of integral models to the choice of $n_s$ is lower than algebraic methods, and the results of the integral method with $n_s = 1$ can be as accurate as the one with $n_s = 3$, especially if the (single) sampling location is selected at the lower edge of the log layer ($h^+ \sim 30$, compare the results of WMLES2 and 4) instead of within the log layer at the lower edge of the overlap region ($h^+ \sim 60$).

### 4.4. WMLES of non-Newtonian fluid flow

The key parameters of the present WMLES models, including WSM closure type, sampling height, sampling cell, and axial grid resolution are analyzed in this section for turbulent non-Newtonian fluid flows. The corresponding WRLES results are considered as the reference solution.

#### 4.4.1. Effects of the WSM model closure

The predicted mean axial velocity profiles with different WSM model closures from table 3 at two sampling heights, i.e., within the log layer near the lower edge of the overlap region ($h^+ \sim 60$) and the upper edge of the buffer layer ($h^+ \sim 30$), are plotted against WRLES prediction in figure 10. For a more quantitative comparison, the relative error of the pipe bulk velocity is also reported in table 4. The result obtained by the simple no-slip boundary condition (WMLES22) substantially deviates from the one of WRLES which conveys the importance of using WSM for coarse LES grid. In addition, comparing figure 10 with figure 8, for the shear-thinning case, the sensitivity of predictions to the WSM type or its parameters is much higher than in the case of the Newtonian fluid flow, especially at higher values of the sampling height ($h^+ \sim 60$). This highlights the importance of the choice of WSM type for TNNF.

More importantly, based on the data reported in figure 10 and table 4, the conventional WSM models developed for Newtonian fluids (WMLES6-11), which are usually used in available flow solvers, all fail to predict the shear-thinning induced drag reduction phenomenon and largely underpredict the bulk velocity. Considering the non-Newtonian rheology in all types of WSM significantly improves the primary flow statistics, i.e., the mean axial velocity. This is confirmed by comparing WMLES13 with 7 at $h^+ = 30$ or WMLES12 with 6 at $h^+ = 60$ for the algebraic-type WSM, WMLES19 with 9 at $h^+ = 30$ or WMLES18 with 8 at $h^+ = 60$ for the integrated-type WSM, and WMLES21 with 11 at $h^+ = 30$ or WMLES20 with 10 at $h^+ = 60$ for the ODE-type WSM in figure 10 and table 4.

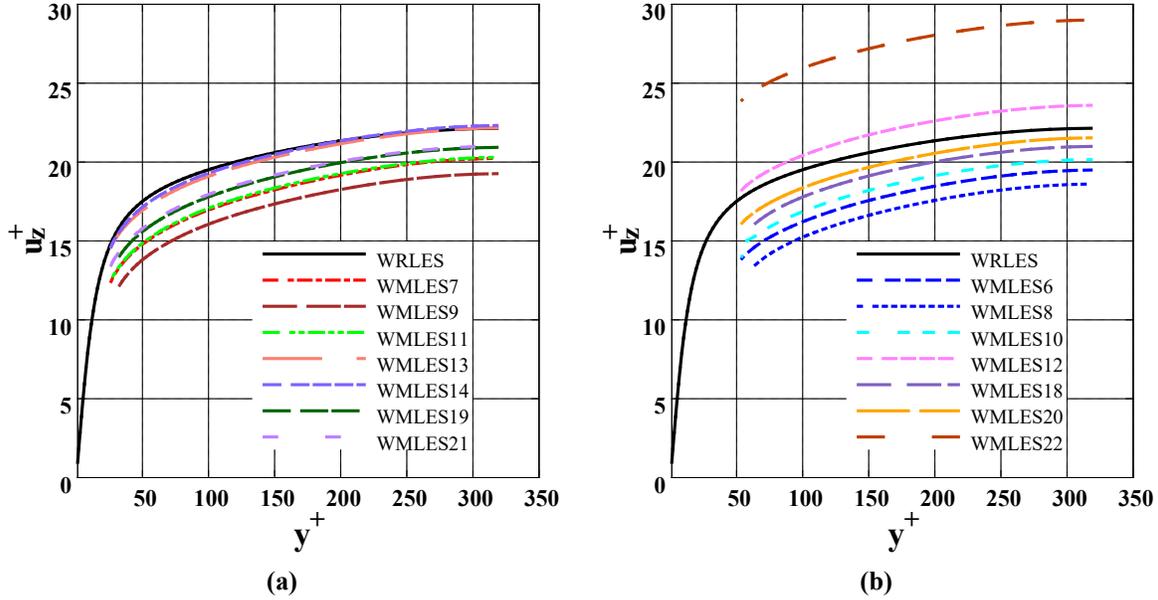

(a)     (b)

Figure 10: The mean axial velocity profile: The comparison of WMLES cases (see table 3) and WRLES predictions for the non-Newtonian test case. The upper bound of the sampling cell is $h^+ = 30$ (a) and $h^+ = 60$ (b).

Table 4: Error (in percentage) in the bulk velocity predicted by WMLES cases (table 3) with respect to the one of WRLES, sorted from the smallest to the largest error. The conventional WSM models are highlighted in grey.

| Case | % | Case | % |
|---|---|---|---|
| **WMLES14** | 0.3 | **WMLES17** | 9.0 |
| **WMLES13** | 1.0 | **WMLES18** | 9.2 |
| **WMLES15** | 1.1 | **WMLES11** | 10.9 |
| **WMLES12** | 4.4 | **WMLES7** | 11.3 |
| **WMLES16** | 5.3 | **WMLES10** | 13.1 |
| **WMLES20** | 5.9 | **WMLES6** | 16.3 |
| **WMLES21** | 6.8 | **WMLES9** | 16.3 |
| **WMLES19** | 7.6 | **WMLES8** | 21.4 |

As already mentioned, integrated-type WSM is superior to the algebraic counterpart with exactly the same parameters. However, in contrast to the Newtonian fluid case study in the previous section, the performance of integrated type with $n_s = 1$ is not close to the algebraic one with $n_s = 3$, and at both sampling heights of $h^+ = 30$ and $h^+ = 60$, using the algebraic WW with $n_s = 3$ produces better results than the integrated WW for the TNNF case. Furthermore, the results of the algebraic model are more accurate than the ODE one, comparing WMLES13 with 21 or WMLES12 with 20. It should be noted that the CPU time of the algebraic NNSS is also higher than the NNODE according to table 3. This is because of the complexity and implicitness of the correlation used for the algebraic NNSS model and the omission of NEE in NNODE for the present simple case studies. To explore the effect of stream-wise grid resolution, WMLES14 can be compared with WMLES13. It is observed that the mean axial velocity profile is not influenced discernably by increasing the axial grid spacing by a factor of 2 in WMLES14, which suggests the relatively low sensitivity of the mean velocity to the stream-wise grid resolution. This survey is compatible with the results reported by Kawai and Larsson [24].

To assess the performance of WMLES in terms of other important flow statistics, the predictions by different WMLESs are compared with the WRLES solution in figure 11 and figure 12. The corresponding errors with respect to the reference solution are also reported in table 5. The error norm has been computed for each parameter, $q$, by:

$$Er = \frac{\sum_{j=1}^{n}(|q_{\text{WRLES},j} - q_{\text{WMLES},j}|)}{\sum_{j=1}^{n}(|q_{\text{WRLES},j}|)} \times 100\% \tag{54}$$

where the summation is performed over all grid points, $j$, across the outer layer, from $0.2\delta^+$ to the center of the pipe.

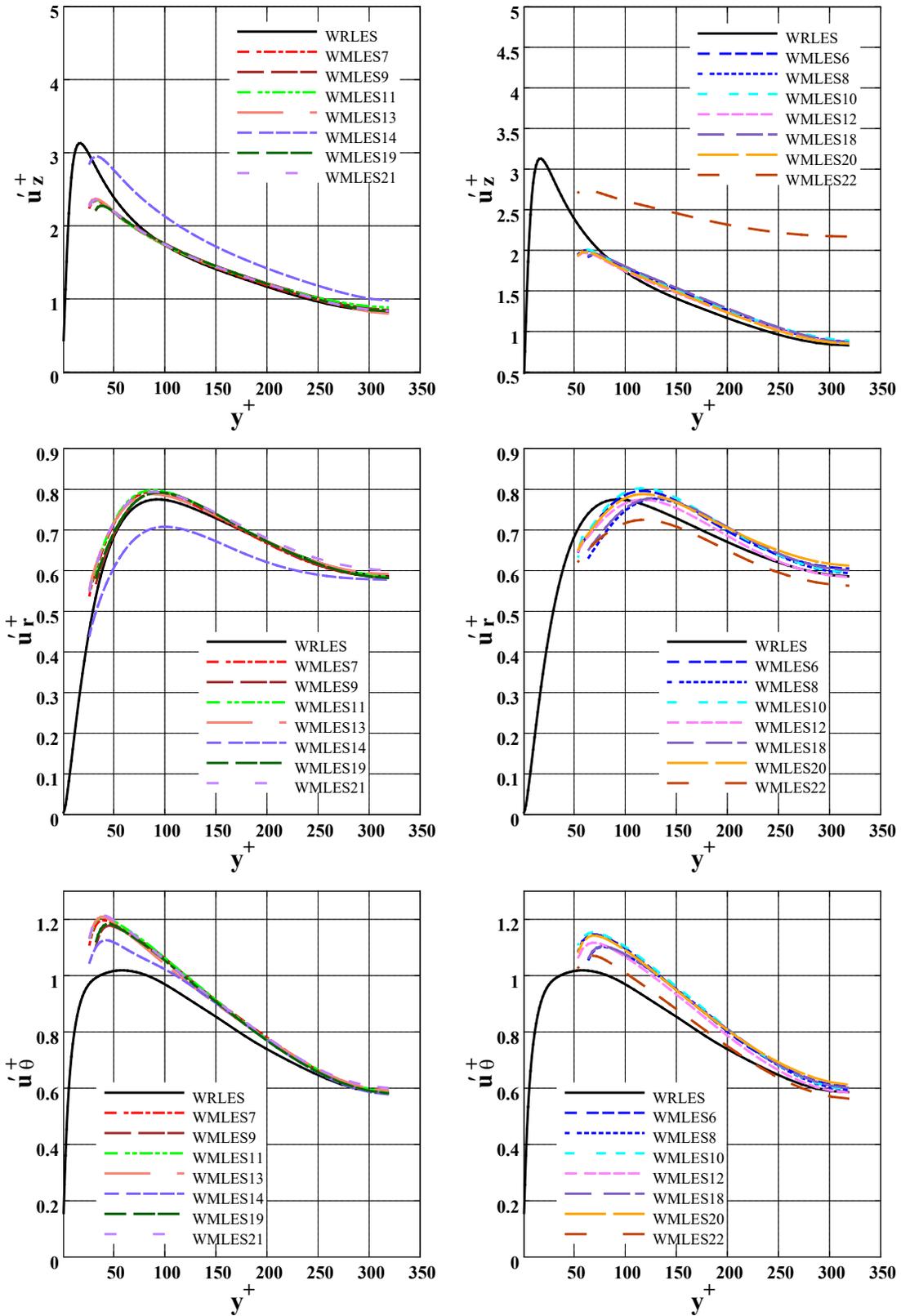

Figure 11: The profile of the resolved RMS velocity, $u_i'^+$, in the axial (top), radial (middle), and azimuthal (bottom) directions: The comparison of WMLES cases (see table 3) and WRLES predictions for the non-Newtonian test case. The upper bound of the sampling cell is $h^+ = 30$ (left) and $h^+ = 60$ (right).

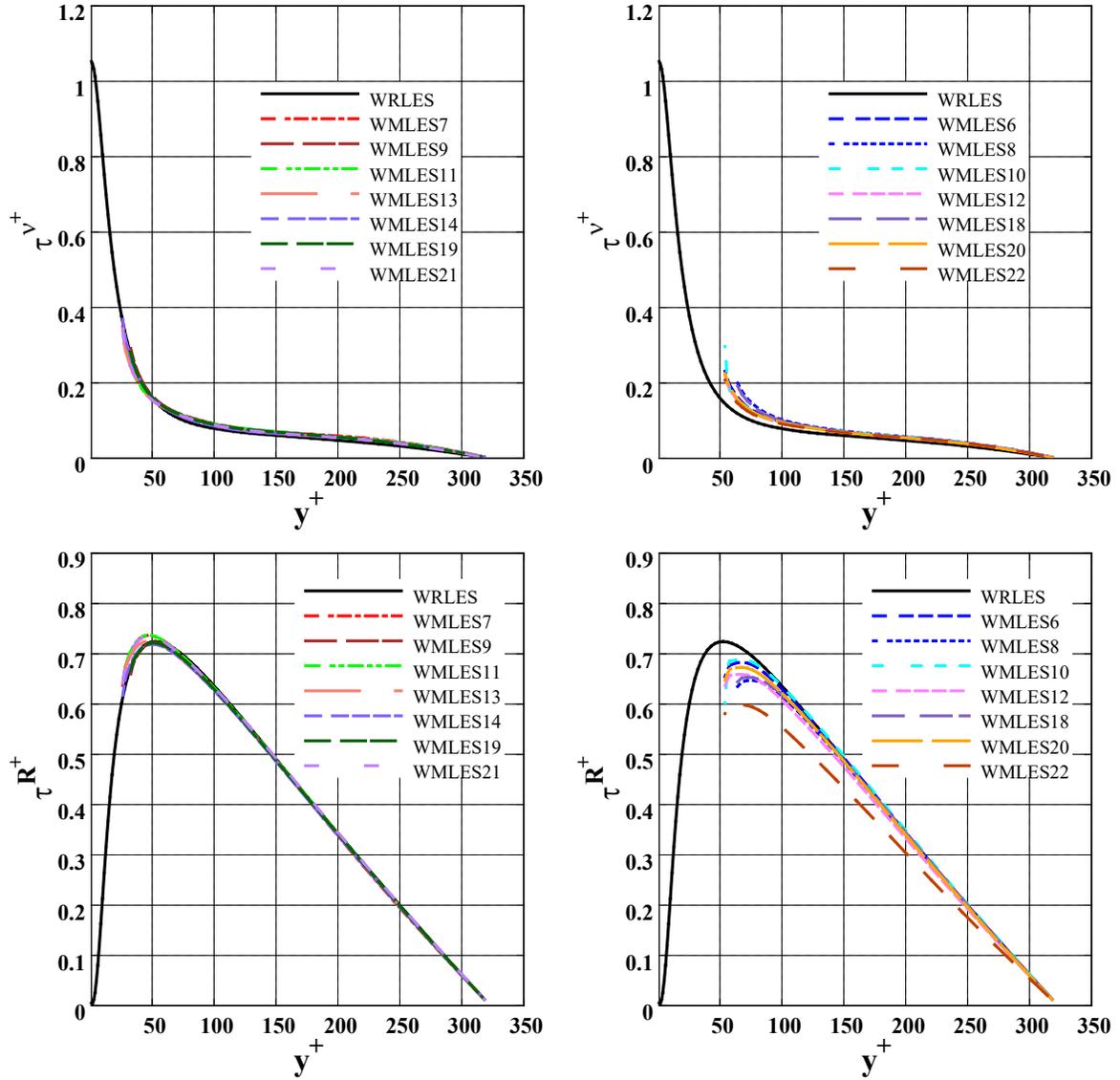

Figure 12: The profile of the mean shear viscous stress, $\tau^{v+} = v^+ \partial u_z^+/\partial y^+$, (top) and shear Reynolds stress, $\tau^{R+} = \langle u_r' u_z' \rangle/u_\tau^2$, (bottom): The comparison of WMLES cases (see table 3) and WRLES predictions for the non-Newtonian test case. The upper bound of the sampling cell is $h^+ = 30$ (left) and $h^+ = 60$ (right).

Using the simple no-slip condition (WMLES22) results in an enormous overprediction of the RMS velocity in the axial direction, $u_z'^+$, while the radial RMS velocity, $u_r'^+$, and shear Reynolds stress, $\tau^{R+}$, are underpredicted compared to other models. This is the same conclusion drawn for the Newtonian case in the previous section.

According to figure 12, the mean shear viscous stress, $\tau^{v+}$, and shear Reynolds stress, $\tau^{R+}$, predicted by all WMLES are in fine agreement with the reference WRLES solution and less sensitive to the choice of WSM, especially with the sampling height of $h^+ = 30$. On the other hand, based on the data of table 5, the conventional Newtonian WSMs (WMLES6-11) have the largest errors in the prediction of mean apparent viscosity, $v^+$, especially with $h^+ = 60$. This corroborates the necessity of accounting for non-Newtonian rheology in WSM. Other models have less than a 3% percent error in predicting the mean apparent viscosity. Note that the error in $v^+$ by the integrated NNSS with $n_s = 1$ and $h^+ = 60$ is greater than 3%. Therefore, in contrast to the Newtonian case, care must be taken when using integrated WSM with sampling only at the wall adjacent cell.

Table 5: The norm of error (in percentage) in different statistics predicted by WMLES with respect to the WRLES reference solution.

| *simulation* | $v^+$ | $\tau^{R^+}$ | $u_z'^+$ | $u_r'^+$ | $u_\theta'^+$ |
|---|---|---|---|---|---|
| **WMLES6** | 3.2 | 1.1 | 5.0 | 3.8 | 10.5 |
| **WMLES7** | 2.6 | 0.8 | 1.8 | 1.5 | 7.4 |
| **WMLES8** | 3.4 | 3.3 | 5.6 | 5.0 | 9.1 |
| **WMLES9** | 2.6 | 0.8 | 2.8 | 1.4 | 7.3 |
| **WMLES10** | 3.5 | 1.2 | 5.5 | 4.1 | 11.2 |
| **WMLES11** | 2.4 | 0.4 | 2.9 | 1.7 | 7.8 |
| **WMLES12** | 2.3 | 4.3 | 4.0 | 2.5 | 7.9 |
| **WMLES13** | 1.8 | 1.1 | 2.3 | 1.1 | 6.5 |
| **WMLES14** | 1.6 | 1.3 | 21.4 | 7.5 | 4.8 |
| **WMLES15** | 1.8 | 2.1 | 1.9 | 1.2 | 6.4 |
| **WMLES16** | 1.6 | 3.3 | 2.6 | 0.7 | 5.9 |
| **WMLES17** | 2.0 | 1.2 | 2.9 | 1.4 | 3.1 |
| **WMLES18** | 3.4 | 2.6 | 6.3 | 5.0 | 9.3 |
| **WMLES19** | 2.2 | 0.8 | 2.3 | 1.3 | 6.8 |
| **WMLES20** | 2.9 | 2.2 | 4.0 | 4.2 | 10.6 |
| **WMLES21** | 2.2 | 0.4 | 2.2 | 2.1 | 7.7 |

Considering figure 11 and table 5, the accuracy of the algebraic NNSS and NNODE WSM types in the prediction of the RMS velocities are close to each other with a slightly better performance of the algebraic one at the higher sampling location. Therefore, the sensitivity of RMS velocity prediction to the type of WSM model is insignificant and the results are close to each other (see figure 11, note that WMLES14 and 22 should be excluded in this figure since the former has a different grid resolution and the latter does not use WSM). The low sensitivity of RMS velocity prediction to the WSM type was also observed by Duprat, et al. [15] for the Newtonian fluid flow.

The comparison of WMLES13 and 14 in figure 12 indicates that the mean viscous stress and shear Reynolds stress profiles do not change discernably by coarsening the axial grid resolution. The low sensitivity of shear Reynolds stress to the stream-wise grid resolution for Newtonian fluid flows has also been reported elsewhere [23, 24]. Comparing WMLES13 and 14 in figure 11, it is revealed that the stream-wise and wall-normal RMS velocities are sensitive to the stream-wise grid resolution. This is not consistent with what was reported by Kawai and Larsson [24]. Besides the difference between the fluid rheology of the present case study and the Newtonian fluid in reference [24], these different observations may be predominantly due to the much finer axial grid resolution considered in their study (the maximum axial grid resolution of $0.084\delta$ versus the value of $0.3925\delta$ in our study). Therefore, their coarsest grid was probably fine enough to resolve the turbulent intensities.

To sum up this section, it is worth mentioning that the best performance in predicting all statistics belongs to WMLES13, i.e., by employing the algebraic NNSS model with the third sampling cell ($n_s = 3$) at height $y^+ = 30$ (at the upper edge of the buffer layer).

*4.4.2. Effects of the sampling height*

According to figure 10 to figure 12, the effect of WSM model and parameters are more pronounced when the sampling location is within the logarithmic layer ($h^+ = 60$) rather than at the edge of the buffer layer ($h^+ = 30$), and the predictions have more disparity in the former situation. From the discussions made in the previous section, the algebraic NNSS shows a better overall performance than the NNODE. To further analyze the sensitivity of the algebraic NNSS to the sampling height keeping the sapling cell ($n_s = 3$) unchanged, the results of WMLES12, 13, and 17 with three values of $h^+ = 60$, 30, and 15 can be considered. In terms of the mean axial velocity, the data in table 4 manifests that the accuracy of the predictions improves notably by decreasing the sampling height from $h^+ = 60$ to $h^+ = 30$ while reducing it further to $h^+ = 15$ not only does not enhance the accuracy but also increases the prediction error again. To justify this observation, we checked the Shenoy-Saini correlation (see section 2.2.4) and found out that it has a larger deviation from the DNS data around $h^+ = 15$. Similarly, in the case of Newtonian fluid flows, Mukha, et al. [16] observed that reducing the height of the sampling cell does not necessarily enhance the results. More or less, the same trend can be seen for other statistics (table 5) which suggests that the best sampling location is $h^+ = 30$ for the algebraic NNSS. NNODE predictions show the lowest sensitivity to the choice of sampling height, however, its accuracy is inferior to the one of the algebraic NNSS in the present case studies.

*4.4.3. Effects of the sampling cell*

It is well established that the first cell off-the-wall is not a proper choice for sampling in algebraic models in Newtonian fluids. We check this point in case of WMLES of TNNF. The test cases are chosen here so as to examine the sampling cell influence on the improved algebraic NNSS WSM performance. figure 13 and figure 14 illustrate the results of algebraic NNSS with different choices of sampling cell, i.e., $n_s = 1, 2,$ and 3 in WMLES16, 15, and 13, respectively, while keeping the other factors, including the grid resolution and $h^+$, unchanged. In terms of the mean axial velocity (figure 13) and bulk velocity (table 4), the best result is obtained with $n_s = 3$; however, there is not much difference between the second and third sampling cells, which was also reported elsewhere [24, 25] for Newtonian fluid flows.

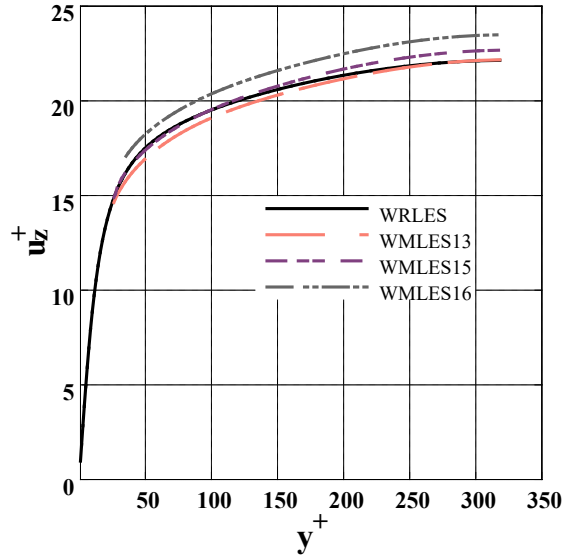

Figure 13: The mean axial velocity profile: The comparison of WMLES16, 15, and 13 (table 3) with different sampling cells of $n_s = 1, 2$, and 3, respectively, and WRLES predictions for the non-Newtonian test case. The upper bound of the sampling cell is $h^+ = 30$.

The mean shear viscous and Reynolds stresses have been exhibited in figure 14a and b, respectively. No distinct sensitivity to the choice of the sampling cell can be observed in the results, at least when the sampling cell resides at the lower edge of the log-layer. Turbulent intensities in the axial and radial direction have been shown in figure 14c and d, respectively. Similarly, there is no significant deviation between the results by changing the sampling cell, even from the first to the second one. Nevertheless, the difference may be more pronounced if a high value of the sampling height is chosen, which has been reported by Kawai and Larsson [24] for a Newtonian fluid flow.

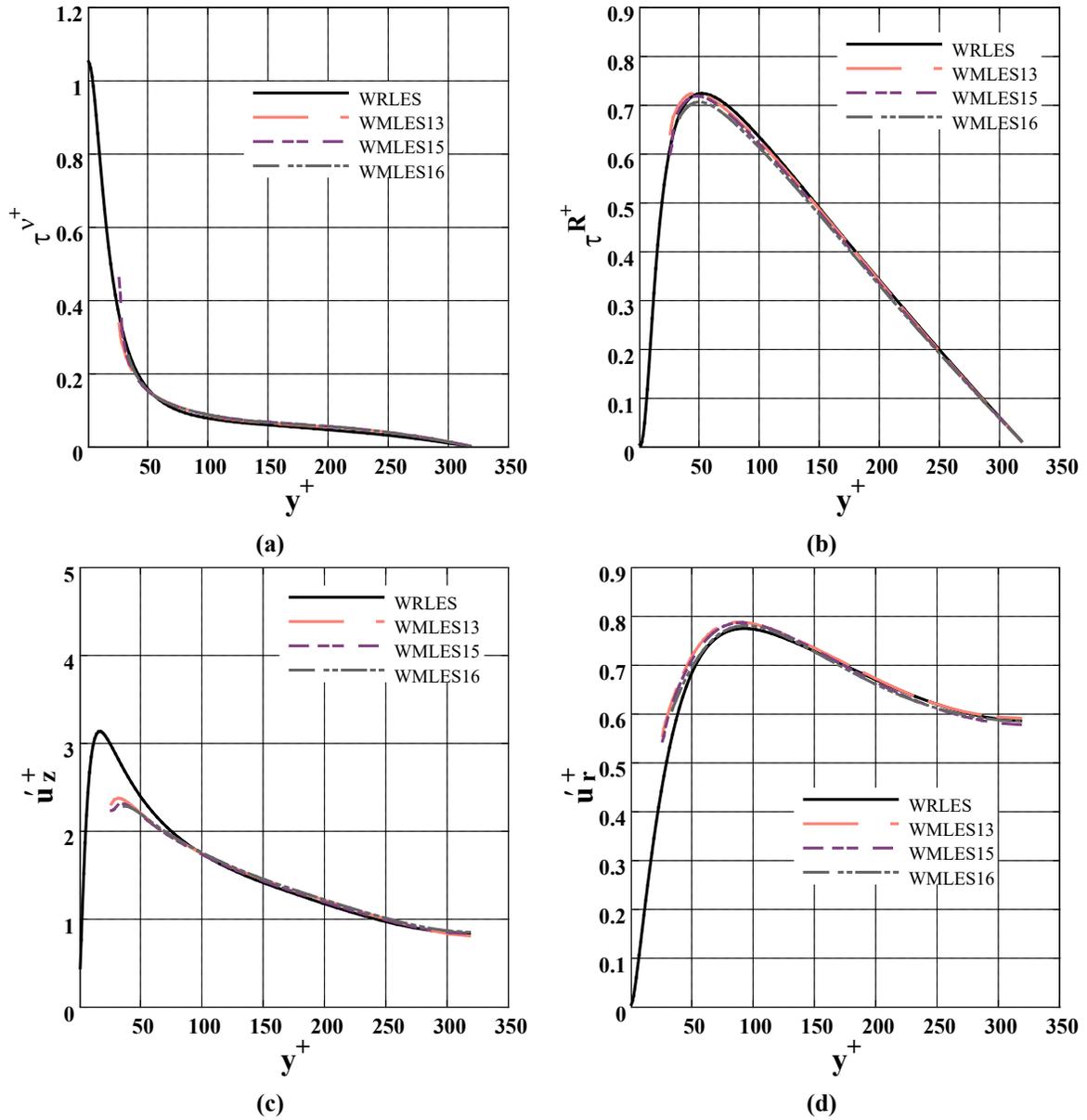

Figure 14: The profile of mean shear viscous stress (a), shear Reynolds stress (b), and resolved RMS velocity in the axial (c) and radial (d) directions: The comparison of WMLES16, 15, and 13 (table 3) with different sampling cells of $n_s = 1$, 2, and 3, respectively, and WRLES predictions for the non-Newtonian test case. The upper bound of the sampling cell is $h^+ = 30$.

*4.4.4. Effects of the viscosity modeling in NNODE*

Based on the results of the previous sections, the algebraic NNSS model had a better performance for turbulent fully-developed pipe flows; however, for more complicated problems involving NEE, the use of ODE-based models may have an advantage. We proved the superiority of the newly proposed NNODE WSM over the conventional ODE one in the previous sections. In this section, we aim to scrutinize the reason behind its better performance by analyzing the effect of different improvements and amendments we introduced in section 2.2.6 to the conventional ODE model to construct the NNODE model. The results of simulations with NNODE WMLES using different assumptions for $\bar{\nu}$ and $\nu_N$ are plotted in figure 15 and figure 16. As it is observed, the influence of modeling these viscosities is not trivial in WMLES, in contrast to a WRLES (see reference [31]), due to the much coarser grid resolution used near the walls in the former case. According to figure 15, the conventional ODE WSM which ignores $\nu_N$ and uses a simple linear model for $\bar{\nu}$ (Eq. (17)) has the largest error in predicting the axial mean velocity profile. When the assumption $\bar{\nu} = \nu_w$ is used in place of the linear assumption along with Eq. (31), i.e., the case referred to as NNODE ($\nu_N = 0$) in figure 15, the error norm of mean velocity predictions reduces from 11.7% to 8%. In addition, by including $\nu_N$ (or $\tau_{ij}^N$ stress) by Eq. (46), i.e., NNODE ($\bar{\nu} = \nu_w$), the mean velocity prediction further improves (the error decreases from 8% to 4.6%). This advocates the important contribution of NNSGS ($\tau_{ij}^N$) with the coarse WMLES grid near the walls. In fact, the solution on the near-wall grid can be seen in the mean sense, like in RANS, due to the coarse computational grid. As a result, NNSGS modeling becomes important, like its counterpart in RANS simulations [31, 35].

To check the effect of $\bar{\nu}$ modeling further, it can be seen in figure 15 that the use of the default model ($\bar{\nu} = \nu_w$ and Eq. (31)), linear model (Eq. (17)), and non-linear model (Eq. (18)) result in 4.6, 6.7, and 8.2% error in predicting the mean axial velocity. Therefore, the non-linear $\bar{\nu}$ model for NNODE WMLES improves the result compared to the linear model, like in RANS simulations [31, 65] but to a lesser extent. However, the presently recommended assumption of $\bar{\nu} = \nu_w$ along with Eq. (31) is found to be the best modeling option for NNODE. This can be justified considering the fact that in addition to the uncertainties of all models, i.e., the modeling errors for the linear and non-linear closures and the assumption of $\bar{v} = v_w$ within the inner layer for the default closure, the linear and non-linear models depend on the calculation of filtered strain rate which carries significant extra errors, especially on the coarse near-wall grid of a WMLES.

Finally, as it can be seen in figure 16, higher order statistics are much less sensitive to the choice of $\bar{v}$ and $v_N$ viscosity modeling in NNODE WSM.

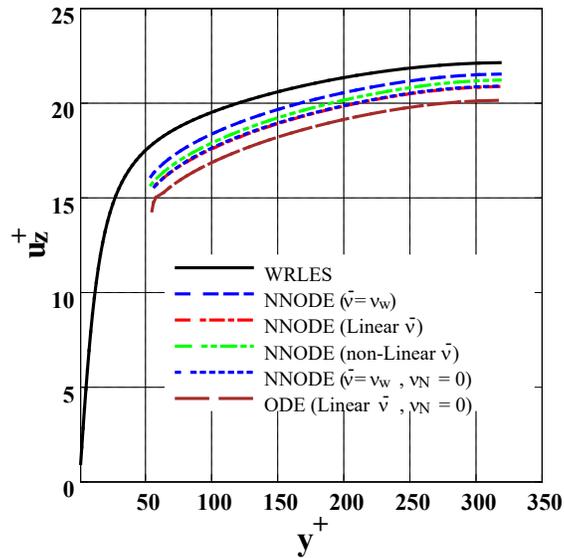

Figure 15: The mean axial velocity profile: The effect of different $\bar{v}$ and $v_N$ modeling approaches, including $\bar{v} = v_w$ (Eq. (31), default), linear $\bar{v}$ (Eq. (17)), non-linear $\bar{v}$ (Eq. (18)), and omitting $v_N$, in NNODE WSM. Other parameters are the same ($n_s = 3$ and $h^+ = 60$).

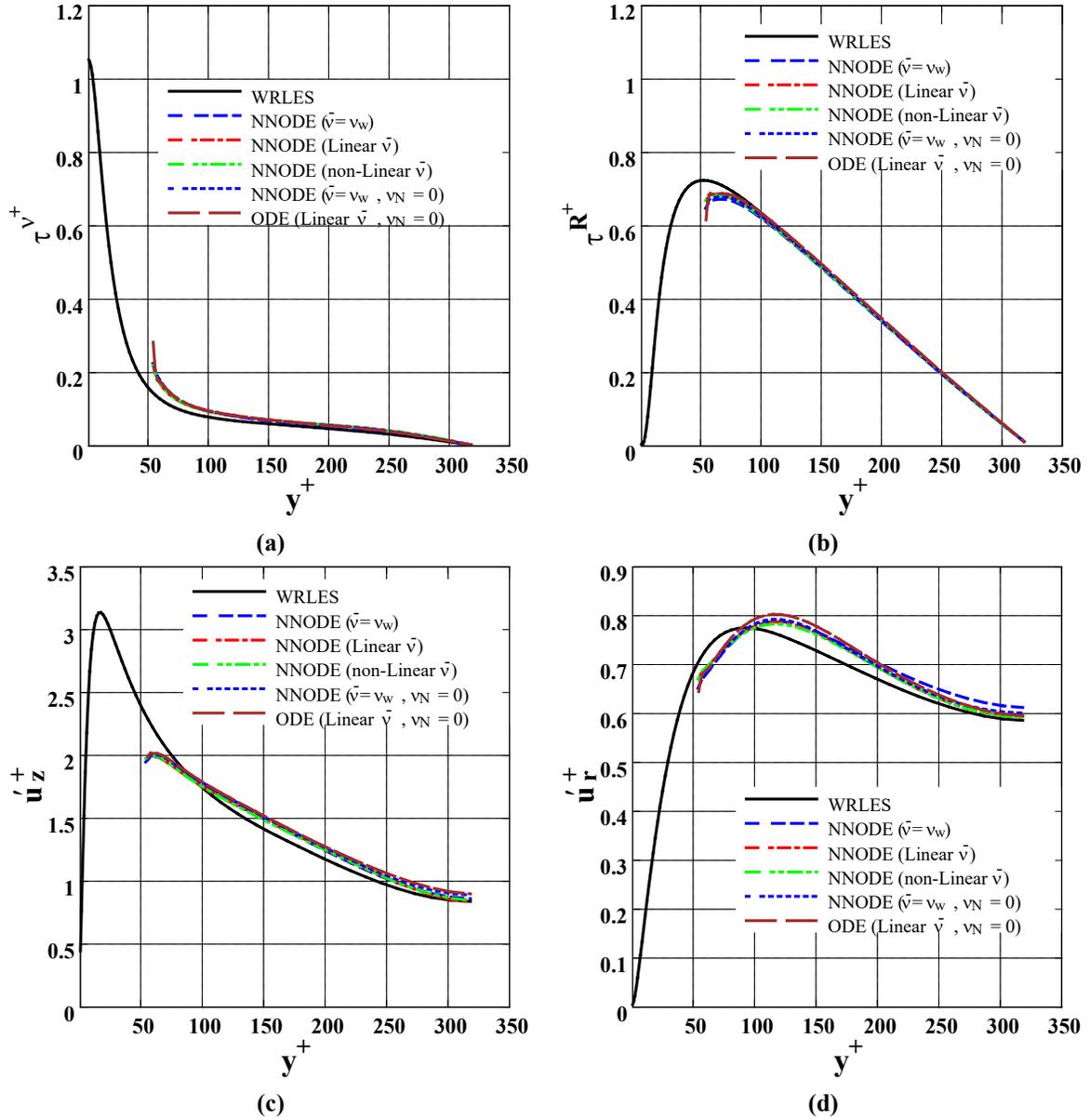

Figure 16: The profile of mean shear viscous stress (a), shear Reynolds stress (b), and resolved RMS velocity in the axial (c) and radial (d) directions. The legend is the same as described in the caption of figure 15.

## 5. Conclusion

In the present research, novel algebraic, integrated, and ODE wall-stress models for WMLES of turbulent power-law fluid flows were proposed and tested successfully in the case of turbulent fully-developed pipe flow of a shear-thinning fluid. For the assessment of new models, WRLES reference solutions were obtained and validated against DNS case studies available in the literature. Then, the effects of WSM type and key parameters, including the sampling height, sampling cell, stream-wise grid resolution, and viscosity closure (in the case of the ODE-type model) on the model performance were carefully investigated.

It was proved that the conventional WSMs developed for Newtonian fluid flows fail to predict the drag reduction phenomenon of shear-thinning fluids and largely underpredict the pipe bulk velocity. The inclusion of NN rheology in WSM significantly improved the prediction of all types of WSM, algebraic, integrated, and ODE, considered in this study. It was observed that, generally, the sensitivity of WMLES to WSM type and parameters is much higher for shear-thinning TNNF than for Newtonian fluid flows, especially as the sampling height increases within the log layer. This highlights the importance of studies to improve WSM closures for TNNF. In addition, it was seen that the WMLES predictions do not necessarily improve by reducing the sampling height. In the case of algebraic NNSS WSM, this observation was attributed to the higher uncertainty of the incorporated NN low-of-the-wall within the buffer layer. Based on the present results and knowing that an integrated WSM almost always performs better than its algebraic counterpart, the best performance for the present pipe flow test cases is obtained via the integrated NNSS WSM sampling at the lower edge of the log layer within the third off-the-wall computational cell. However, the proposed NNODE WSM probably offers better results for more complicated problems involving NEE. This model also showed less sensitivity to the model parameters than the others. Different closures for the mean apparent and NNSGS viscosities were assessed for this model. Finally, the proposed NNODE model improved the mean velocity predictions by 7.1% compared to the conventional ODE model.

The present models can be helpful for WMLES of complex practical TNNF problems, and their evaluation in more complicated case studies would be an interesting topic for future studies.

**Funding.** This research received no specific grant from any funding agency, commercial or not-for-profit sectors.

**Declaration of interests.** The authors report no conflict of interest.

**Appendix A**

Different quality criteria have been proposed in the literature to check whether an LES solution can be deemed well-resolved. One of these criteria is the well-known Pope's criterion [48] which suggests that for a well-resolved LES, the filter width, i.e., the grid size in the case of the implicit filter, should be chosen such that at least 80% of the turbulent kinetic energy is resolved locally. The ratio of the resolved to total turbulent kinetic energy against the distance from the wall is shown in figure 17a. As observed, more than 90% of turbulent kinetic energy was resolved except where $y^+ < 2$, which is the inner part of the laminar viscous sublayer and is attributed to the uncertainty of the SGS kinetic energy estimation by Eq. (51) very close to the walls [31], rather than an issue with the WRLES solution. The other criterion used here is based on the ratio of mean SGS viscosity to the mean apparent viscosity. Celik, et al. [66] suggested that this ratio should be less than 20, which is associated with the LES quality index of 0.8, for a well-resolved LES. For the present WRLES, this ratio is illustrated in figure 17b. As can be observed, this ratio is far below this threshold, indicating a high-quality LES simulation.

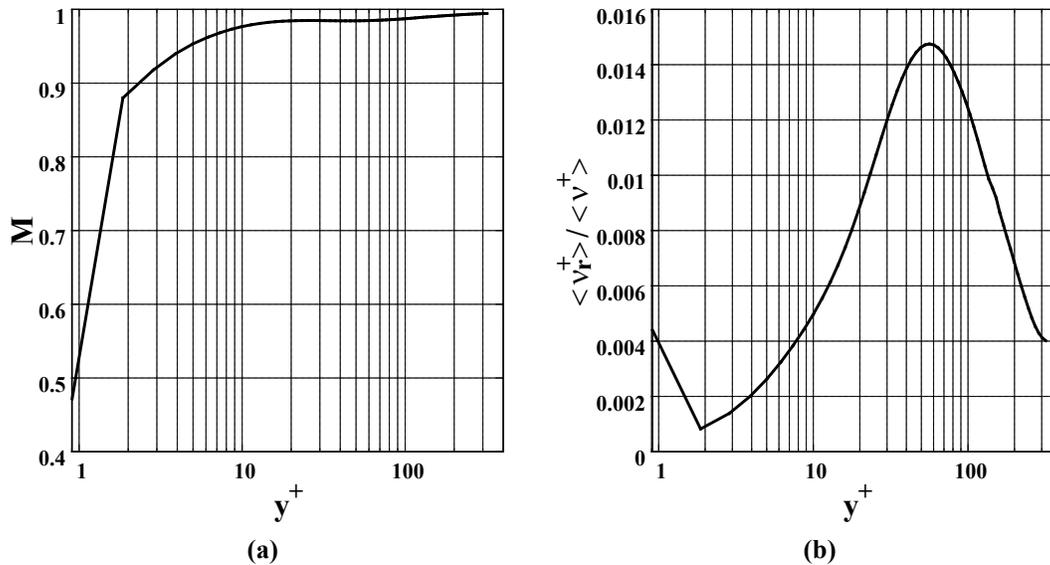

Figure 17: Pope criterion: The ratio of the resolved to total turbulent kinetic energy (a) and the ratio of the mean SGS viscosity to mean apparent viscosity (b) versus the distance from the wall for the non-Newtonian WRLES solution on the medium (main) grid.